\documentclass[useAMS,usenatbib]{mn2e}
\usepackage{times}
\usepackage{rotating}
\usepackage{amsmath}
\usepackage{amssymb}
\usepackage[dvipsnames]{xcolor}
\usepackage{graphicx}
\usepackage{graphics}
\usepackage{hyperref}
\hypersetup{
     colorlinks   = true,
     citecolor    = blue
}

\def\apj{{ ApJ}}
\def\apjl{{ApJL}}

\def\aap{{ A\&A}}

\def\mnras{{ MNRAS}}

\def\nat {{ Nature}}
\def\pasj{{ PASJ}}

\def\prd{{ Phys. Rev. D}}

\def\pasa{{PASA}}
\def\aapr{{A\&ARv}}

\long\def\symbolfootnote[#1]#2{\begingroup%
\def\thefootnote{\fnsymbol{footnote}}\footnote[#1]{#2}\endgroup}
\newcommand{\gae}{\lower 2pt \hbox{$\, \buildrel {\scriptstyle >}\over {\scriptstyle
\sim}\,$}}
\newcommand{\lae}{\lower 2pt \hbox{$\, \buildrel {\scriptstyle <}\over {\scriptstyle
\sim}\,$}}

\def\d{\mathrm{d}}
\def\flu{\mathcal{F}}
\def\mr{\mathrm}
\def\pp{{\prime\prime}}
\def\nGJ{\mbox{\large n}_{_{GJ}}}

\def\r5{\rho_{_5}}

\begin{document}

\title[FRB radiation mechanism]
{Fast radio burst source properties and curvature radiation model}

\author[Kumar, Lu \& Bhattacharya]{Pawan Kumar$^{1}$\thanks{E-mail:
    pk@astro.as.utexas.edu,  
wenbinlu@astro.as.utexas.edu}, Wenbin Lu$^1$ and Mukul Bhattacharya$^{2}$  
\\ $^{1}$Department of Astronomy, University of Texas at Austin, Austin, 
 TX 78712, USA\\
$^{2}$Department of Physics, University of Texas at Austin, Austin, 
 TX 78712, USA}

% \pagerange{\pageref{000}--\pageref{000}} \pubyear{2017}

\maketitle

\begin{abstract}
We use the observed properties of fast radio bursts (FRBs) and a number of
general physical considerations to provide a broad-brush model for the
physical properties of FRB sources and the radiation mechanism. We
show that the  
magnetic field in the source region should be at least 10$^{14}$ Gauss. This
strong field is required to ensure that the electrons have sufficiently high 
ground state Landau energy so that particle collisions, instabilities, 
and strong electro-magnetic fields associated with 
the FRB radiation do not perturb electrons' motion in the direction
transverse to the magnetic field and destroy their coherent motion; coherence is required by the high observed brightness temperature of FRB radiation. 
The electric field in the source region required to sustain particle motion
for a wave period is estimated to be of order 10$^{11}$ esu. These
requirements suggest that FRBs are produced near the surface of
magnetars perhaps via forced  
reconnection of magnetic fields to produce episodic, repeated, outbursts. 
The beaming-corrected energy release in these bursts is   
estimated to be about $10^{36}$ ergs, whereas the total energy in the 
magnetic field is at least $\sim10^{45}$ ergs. We provide a number of
predictions for this model which can be tested by future
observations. One of which is that short duration FRB-like bursts
should exist at much higher frequencies, possibly up to optical.
\end{abstract}


\section{Introduction}

Fast radio bursts (FRBs) are Jansky-level millisecond duration 
transient events at $\sim$ GHz frequencies of unknown physical origin
discovered in pulsar surveys \citep{2007Sci...318..777L,
  2013Sci...341...53T}. They have an all-sky rate of $\sim10^3$ to
$10^4\rm\ day^{-1}$ above fluence $\sim 1\rm\ Jy\ ms$
\citep{2013Sci...341...53T, 2015MNRAS.447.2852K, 2016MNRAS.455.2207R,
  2016MNRAS.tmpL..49C}. Very recently, the location of one of these
bursts FRB 121102, a repeater, has been determined to
an accuracy of $\sim0.1^\pp$ by interferometry with the Jansky Very Large  
Array \citep{2016Natur.531..202S, 2016ApJ...833..177S,
  2017arXiv170101098C}. This FRB is found to be 
associated with a dwarf star-forming host galaxy at redshift z =
0.19273 \citep{2017ApJ...834L...7T}. The FRB is also associated with a
persistent radio source \citep{2017arXiv170101098C}, and their
separation is further pinned down to $\lesssim0.01^\pp$ ($\lesssim
40\rm\ pc$ in physical distance) by the European VLBI Network
\citep{2017ApJ...834L...8M}.

It is not surprising that the distance of FRB 121102 is $\sim 1$ Gpc 
considering that many observational evidences had already suggested
that FRBs are at cosmological distances \citep[see][for a brief 
review]{2016MPLA...3130013K}. First of all, their dispersion measures (the 
column density of free electrons along the line of sight $DM=\int n_{\rm e} 
\d l\sim 10^3\rm\ pc\ cm^{-3}$) are much larger than the contribution
from the interstellar medium (ISM) in the Milky Way by roughly an 
order of magnitude\footnote{A list of all reported FRBs and their
  properties can be found at
  \href{http://www.astronomy.swin.edu.au/pulsar/frbcat/}{http://www.astronomy.swin.edu.au/pulsar/frbcat/}
  \citep{2016PASA...33...45P}.}. Second, if the DM is mostly
contributed by an ionized nebula with electron temperature $T_{\rm
  e}\sim 10^4\rm\ K$, then in order to avoid free-free absorption of GHz 
waves the size of the nebula must be larger than $\sim 0.01\rm\ pc$
\citep{2014ApJ...785L..26L}. Moreover, the strength of H$\alpha$ and UV
continuum flux limits combined with the free-free absorption argument
suggest that at least the ``Lorimer burst'' FRB 010724 is at a distance
$d\gtrsim 1\rm\ Mpc$ \citep{2014ApJ...797...70K}.

Confirmation of the cosmological origin means that FRBs are very
energetic events. If the FRB sources are isotropic (the effect of 
anisotropy will be included later on), then the energy release is
\begin{equation}
  \label{eq:8}
  E_{\rm iso} = (1.3\times10^{40} \ \mr{erg}) \frac{\flu}{\rm Jy\ ms}
  d_{28}^2 \Delta \nu_9, 
\end{equation}
where $\flu$ is the fluence, $d = 10^{28}d_{28}\rm\ cm$ is the
luminosity distance and $\Delta\nu = \Delta\nu_9 \rm\ GHz$ is the
width of the FRB spectrum. If FRB radiation is incoherent, for burst
duration $\delta t = \delta t_{-3}\rm\ ms$, the area of the emitting 
region in the transverse direction --- for a source with sub-relativistic 
speed --- is $A< \pi (c\delta t)^2$, so the brightness temperature is
\begin{equation}
  T_{\rm B} = \frac{S_{\nu} d_{\rm A}^2 c^2}{2 A\nu^2 k_B} > (1\times10^{36}{\rm\:
  K}) \frac{S_{\nu}}{{\rm Jy}}d_{\rm A, 28}^2\delta t_{-3}^{-2}\nu_9^{-2},
  \label{Tb}
\end{equation}
where $S_{\nu}$ is the peak specific flux, $d_{A}$ is the angular
diameter distance ($d_{A,28}\equiv d_A/10^{28}$cm), $\nu = \nu_9 \rm\ GHz$ is 
the lab-frame frequency, $c$ is speed of light, and $k_B$ is the Boltzmann 
constant; we use the convenient notation $X_n\equiv X/10^n$ throughout
the paper. Note that if the source is moving
toward the Earth at Lorentz factor $\gamma$, then $T_{\rm B}^\prime$ (the
comoving frame temperature) is smaller by a factor of $\gamma^3$
than estimated in equation (\ref{Tb}). For any reasonable Lorentz factor, 
the brightness temperature exceeds the Compton catastrophe limit 
of $T'_B \sim 10^{12}$ K. This means that FRB radiation must be 
coherent and the emitting particles relativistic, as pointed out 
by \citet{2014PhRvD..89j3009K}.

Regarding the nature of FRBs, the major unknowns are their
progenitors and radiation mechanism. Many progenitor models have been
proposed, including: collapsing neutron stars
\citep[][]{2014A&A...562A.137F, 
  2014ApJ...780L..21Z}, neutron star or white dwarf mergers
\citep{2012ApJ...755...80P, 2013ApJ...776L..39K,
  2013PASJ...65L..12T}, magnetar bursts\footnote{It should be noted that
\citet{2016ApJ...827...59T} failed to observe a FRB during
SGR1806-20's big outburst.} \citep{2010vaoa.conf..129P,
  2014MNRAS.442L...9L, 2015ApJ...807..179P, 2016ApJ...826..226K,
  2016MNRAS.461.1498M}, 
supergiant pulses from young pulsars \citep{2016MNRAS.458L..19C,
  2016MNRAS.457..232C, 2016ApJ...818...19K, 2016arXiv160302891L,
  2016arXiv161101243K}, (Galactic)   
flaring stars \citep{2014MNRAS.439L..46L}, relativistic jets running
into clouds \citep{2016PhRvD....93...3001}, asteroids colliding with neutron stars 
\citep{2015ApJ...809...24G, 2016arXiv160308207D}, close neutron
star-white dwarf binaries \citep{2016ApJ...823L..28G}, lightning in
the neutron star magnetosphere \citep{2017arXiv170202161K},  and
plasma stream sweeping across the neutron star magnetosphere
\citep{2017arXiv170104094Z}. These works
are only based 
on considerations of energetics and timescales, and the authors simply
{\it assumed} that (a fraction of) the free energy available in the
system is radiated away at GHz frequencies by coherent charge 
``patches'' via e.g. curvature or synchrotron processes. 
However, the properties of the emitting particles/plasma and
conditions for coherent emission have not been discussed in
detail and will be the focus of this paper.

\citet{2016MNRAS.457..232C} used
coherent curvature radiation to explain both FRBs and the MJy shot
pulses from the Crab pulsar, but the formation and stability of the
relativistic near neutral coherent patches\footnote{To explain the MJy
  shot pulses from the Crab pulsar, the coherent patch has fractional 
charge (net charge divided by the total number of particles)
$\lesssim 10^{-7}q$; particles of positive and negative charges are
moving in the same direction. The basic reason for requiring an
extremely small fractional charge is that the energy of
FRB electromagnetic waves comes from particles' initial kinetic
energy in the model of \citet{2016MNRAS.457..232C}. The cooling time
is much shorter than 1 ns (see \S3) unless an amount of net charge
$q$ is associated with kinetic energy $\gg \gamma m_e
c^2$. Another reason for quasi-neutrality is the electrostatic 
repulsion.} in their model were not discussed. \citet{2017MNRAS.465L..30G}
derived the general conditions for synchrotron maser emission; GHz
synchrotron emission requires weak (and ordered) magnetic field 
($\lesssim 10^3/\gamma^2$ G for electrons and $\lesssim 10^6/\gamma^2$
G for protons) which can be found at distances $\gtrsim 10^{8\mbox{-}9}\rm\ 
cm$ from a neutron star for a dipole field configuration $B\propto r^{-3}$; 
where $\gamma$ is the Lorentz factor of particles perpendicular to 
the B-field.  \citet{2017MNRAS.465L..30G}
pointed out that the energy requirement for FRBs may be hard to satisfy
for the synchrotron maser model if FRBs are at cosmological
distances. In an earlier paper, 
\citet{2014MNRAS.442L...9L} proposed that synchrotron maser emission
may be produced when a magneto-hydrodynamic wave generated in 
a magnetar giant flare interacts with the surrounding pulsar wind 
nebula at a distance of $\gtrsim 10^{16}\rm\ cm$ (here $B\propto
r^{-1}$). The common problem of \citet{2017MNRAS.465L..30G} and
\citet{2014MNRAS.442L...9L} models is that they did not discuss how the
particle distribution (population inversion) for efficient maser
emission can be achieved.

Most FRBs models are based on neutron stars (NS), which could naturally
explain their short durations, large energy requirement, ordered
magnetic field (hereafter B-field) needed for coherent emission, and
repetitions with intervals between $\sim10^2$ to $\gtrsim10^7$ seconds
\citep[the only repeater so far is FRB 121102, but others may
also be repeating, see][]{2016MNRAS.461L.122L}. In
this paper, we use general physical considerations to show that the
B-field in the FRB source plasma should be $\gtrsim 10^{14}\rm\
G$ and hence FRBs are most likely to be produced in the
magnetosphere of a NS or stellar-mass black hole. We also note that
the persistent radio source near FRB 121102 is consistent with being a
supernova remnant (SNR) energized by a young 
NS/magnetar \citep{2017arXiv170102370M, 2017arXiv170104815K},
although the possibility of an active galactic nuclei (AGN) cannot be
ruled out \citep{2017ApJ...834L...7T}.

Much of the work presented in this paper was done in the summer of 2016, 
but the writing up of the paper took a long time due to some pressing matter. 
We have made every effort to cite papers related to this work that have 
been published in the meantime in this fast developing field.

In \S2, we provide estimates
  for the size of the source 
region, and the strength of the electric field associated with the
FRB radiation at the source. In \S3, we show that coherent curvature
radiation in neutron star magnetosphere can explain the FRB
properties, 
and we also provide a number of arguments to show that the B-field 
strength should be $\gae 10^{14}$ G. Some predictions 
of this model are discussed in \S4, and a summary of the main results
are provided in \S5.
% Throughout the paper, we use CGS units with
% convention $G = 10^xG_x$.

\section{General considerations}\label{sec:general}

We provide a broad-brush picture of the FRB source in this section 
based on general physics considerations.

Considering that the FRB radiation is coherent, the size of
the source along the line of sight cannot be larger 
than the wavelength ($\lambda$) of the radiation we observe. If the
source is moving toward us with a Lorentz factor 
$\gamma$, the frequency in the source frame is smaller than in the
lab frame by a factor $\gamma$, and the transverse size corresponding
to the wavelength is $\gamma\lambda$.  Since particle velocities
  are $(1 - \gamma^{-2})c$, separation of $\lambda$ along the line of
  sight can be maintained over a lab-frame time $\sim
  \gamma^2\lambda/c$. 
The transverse source size is often taken to be
$\ell_t = \gamma\lambda$. However, this need not be the case. The
transverse size can be larger than $\gamma\lambda$ by factor 
$\eta^{1/2}$ and still coherence can be maintained; $\eta\equiv (\ell_t/\gamma
\lambda)^2$. This is because the 
time delay between the arrival of photons at the observer from the
opposite ends of the source in the transverse direction
$\eta(\gamma\lambda)^2/(c\,\min\{d_A, d_t\})$ is smaller than $\nu^{-1}
= \lambda/c$ provided that $\eta < [\min\{d_A,
d_t\}/(\gamma^2\lambda)]$; where $d_A$ is the angular diameter
distance to the source, and $d_t$ 
is the distance between the source and the {\it trigger point} --- located
behind the source --- from which the signal originates and propagates
outward and triggers different points in the source to start radiating.
For FRBs at distances $d_A\sim 10^{28}\rm\ cm$, the requirement on
$\eta$ is
\begin{equation}
  \label{eq:16}
  \eta^{1/2} \lae \min\{d_{t,6}^{1/2}, 10^{11} d_{A,28}^{1/2}\}
\nu_9^{1/2}\gamma_2^{-1}.
\end{equation}
Moreover, the shape of the coherent patch 
in the transverse direction is unknown. In the following, we take the 
area of the coherent patch in the transverse direction to be
\begin{equation}
  \label{eq:30}
  A_{coh} \equiv \eta \gamma^2\lambda^2,
\end{equation}
where $\eta=\eta_x\eta_y$ contains two multiplication factors $\eta_x$
and $\eta_y$ corresponding to the directions of the two principal axes in
the transverse plane.

Thus, the maximum coherent volume in the lab frame is
\begin{equation}
   {V}_{coh} \sim \eta\gamma^2\lambda^3, 
  \label{Vcoh}
\end{equation}
and in the comoving frame
\begin{equation}
{V}'_{coh} \sim \eta\gamma^3 \lambda^3.
\end{equation}
This coherent patch contributes to the FRB radiation for a time duration of
order $\nu^{-1} = \lambda/c$ in the observer's frame, during which the
source has traveled a distance $\sim\gamma^2 \lambda$ (in the lab frame)
toward the observer. The patch turns off after this time and another patch 
lights up so that the intrinsic FRB duration is much longer than
$\nu^{-1}$. There could
  be more than one coherent patches ($N_{patch}$) adding up
  incoherently, and a continuous plasma flow could produce coherent
  radiation for a duration longer than $\nu^{-1}$; two sources that are
  separated by a distance larger than the size of a coherent zone are
  treated as two independent patches, and even within a coherent zone
  we can have multiple patches if the radiation they produce is not in
  phase. If the same source produces coherent radiation continuously
  for a time duration longer than $\nu^{-1}$ (in observer's frame), then
  we tag it as a different source after time interval $\nu^{-1}$ 
  and yet another source after $2\nu^{-1}$ and so on.

The total isotropic equivalent 
of energy release for FRBs is $E_{iso} \sim 10^{40}$ erg and
the isotropic equivalent luminosity is $L_{iso}\sim 10^{43}$ erg s$^{-1}$ 
(for a cosmological distance of $\sim 10^{28}$ cm). In
the far field (Fraunhofer diffraction) limit, the solid angle
within which electromagnetic (EM) waves add up coherently is\footnote{
For a coherent patch with rectangular shape in the transverse
direction and area $A_{coh} = \eta_x\eta_y \gamma^2\lambda^2$, the
beaming solid angle $\Omega_F \sim \gamma^{-2}
\mathrm{min}(\eta_x^{-1}, 1)\times \mathrm{min}(\eta_y^{-1},
1)$. Equation (\ref{eq:15}) is only correct in the limit $\eta_x>1$ and
$\eta_y>1$. For smaller $\eta$'s, the energy density associated with
electromagnetic waves is larger than our current, conservative,
estimate, because more energy is produced in a smaller coherent
volume.} 
\begin{equation}
  \label{eq:15}
  \Omega_F \sim \pi (\eta\gamma^2)^{-1}.
\end{equation}
If there are $N_{patch}$ distinct coherent patches, each of volume
$\sim{V}_{coh}$, contributing to the total observed FRB flux 
{\it at any given time}, the beaming corrected total energy is
\begin{equation}
  \label{eq:29}
  E_{frb} \sim \frac{E_{iso}\Omega_F}{4\pi N_{patch}} \sim
  (3\times10^{35} {\rm\: erg}) 
  E_{iso,40} N_{patch}^{-1} \eta^{-1} \gamma_2^{-2}.
\end{equation}
The average lab-frame energy density in EM waves in the radiating
plasma is
\begin{equation}
   \epsilon_{_{EM}} \sim {L_{iso} (\Omega_{F}/4\pi)\over
     c N_{patch} (V_{coh}/\lambda)} \sim {L_{iso} \over 4\,c
     N_{patch}\eta^2\gamma^4\lambda^2 }, 
\end{equation}
which is of order 10$^{21} L_{iso,43} N_{patch}^{-1}\eta^{-2} 
 \gamma_2^{-4} \nu_9^{2}$ erg cm$^{-3}$. 
Therefore, the electric and magnetic field strengths associated with this EM wave
energy density are
\begin{equation}
  E_{\perp EM} = B_{\perp EM} \sim (4 \pi \epsilon_{_{EM}})^{1/2} 
    \sim ({10^{11}\rm\ esu})\; {L_{iso,43}^{1/2}\nu_9 \over N_{patch}^{1/2}\eta\, 
   \gamma_2^2},
   \label{E_em}
\end{equation}
where the B-field strength is in Gauss. The electric field is
  very strong in the sense that $q E_{\perp EM} \lambda/(m_e c^2) \sim
  10^8$. Thus,  unless this field is almost exactly perpendicular to
  the local magnetic field it will accelerate electrons to highly
  relativistic speeds and drain the energy out of the FRB radiation.

Since all the photons from a patch must be traveling in nearly the same 
direction and in phase in order that their fields add coherently, we therefore 
infer that the electric and magnetic fields associated with the
observed FRB radiation calculated above is in a direction roughly
perpendicular to the patch's velocity vector.

\section{FRB source properties: coherent curvature radiation}
\label{curvature-rad}

Guided by general considerations of the last section that the
B-field in the region where the observed radio photons 
from FRBs are produced is large, we restrict ourselves to 
the magnetosphere of a neutron star or a stellar-mass black hole.
The curvature radiation --- radiation produced when charged particles 
stream along curved B-field lines --- is an efficient way in 
these conditions to produce large coherent radiation. In this section,
we describe a detailed curvature radiation model that can reproduce
observed FRB properties without any fine tuning of parameters.

\begin{figure}
  \centering
\includegraphics[width = 0.45 \textwidth,
  height=0.22\textheight]{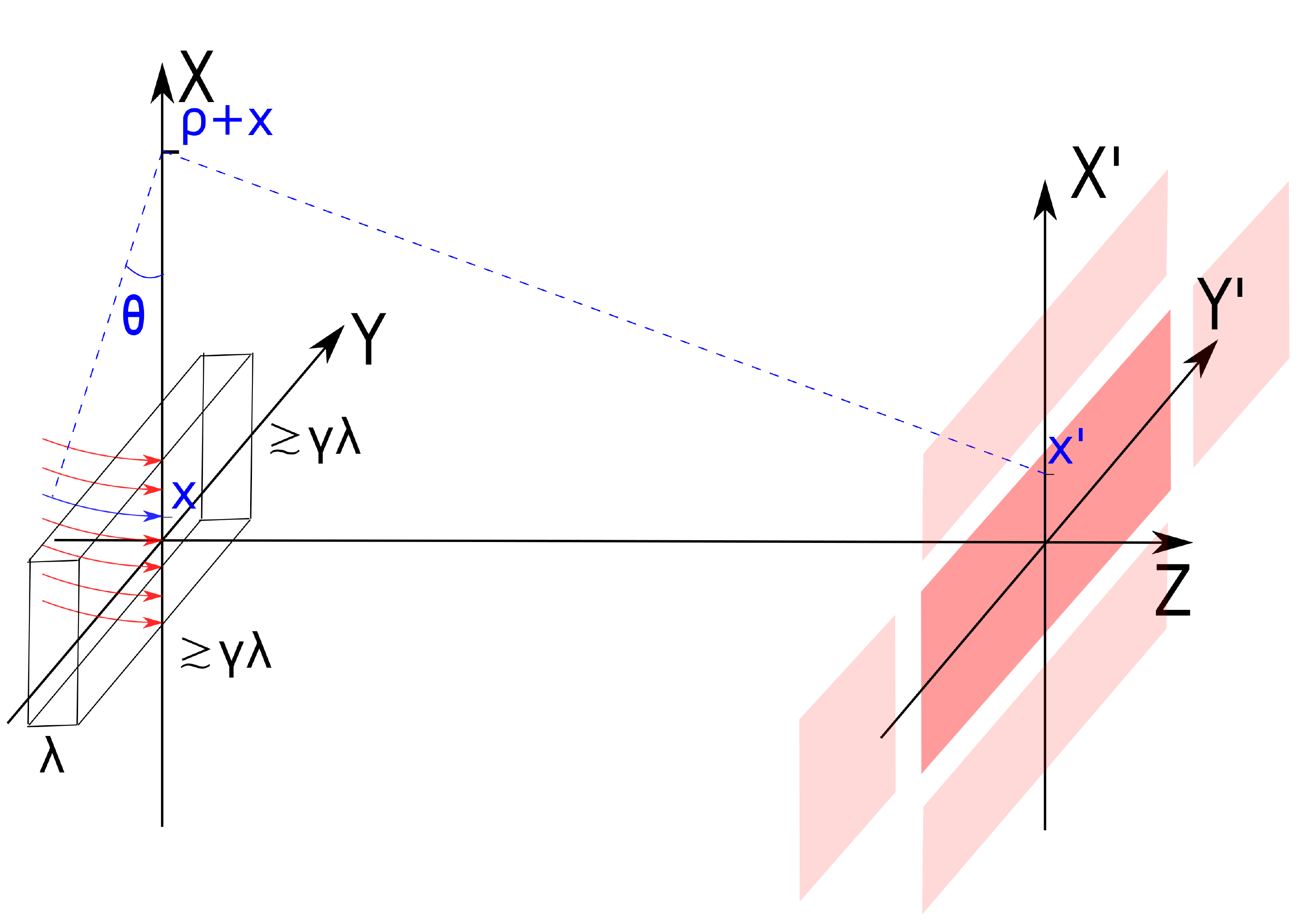}
\caption{One slice of the plasma in the plane perpendicular to the
  line of sight. The
  distance between the source plane (X-Y) and image plane
  (X$^\prime$-Y$^\prime$) is $d$. The coherent emitting volume has
  dimension $\Delta X\gtrsim \gamma\lambda$, $\Delta Y \gtrsim \gamma
  \lambda$ and $\Delta Z \sim \lambda$. The transverse area of the
  coherent patch is $A_{coh}= \eta \gamma^2\lambda^2$.
  In the far field (Fraunhofer diffraction) limit, the first fringe
  has area $\sim d^2(\eta\gamma^2)^{-1}$ and the observer is in this fringe.
}\label{fig:volume}
\end{figure}

Let us consider an electron moving with Lorentz factor $\gamma$ along 
a B-field line of local curvature radius $\rho$. Due to its
acceleration along the B-field curvature, the electron produces
EM waves at frequency
\begin{equation}
   \nu \approx {c \gamma^3 \over 2\pi \rho}, 
   \label{nu_R}
\end{equation}
and the power radiated is
\begin{equation}
  p_e \approx {2 q^2 \gamma^4 c\over 3 \rho^2} \sim (2\times10^{-13} {\rm \:
      erg\,s^{-1}})\; \nu_{9}^{4/3} \r5^{-2/3}.
    \label{pe}
\end{equation}
The second expression for $p_e$ is obtained by expressing $\gamma$ in
terms of $\nu$ using equation (\ref{nu_R}), i.e.
\begin{equation}
    \gamma \approx 28\, \nu_9^{1/3} \r5^{1/3}.
    \label{gam}
\end{equation}

The electron's radiation is beamed in the forward direction
within a cone of angle $\gamma^{-1}$, 
and the pulse time in the observer frame is smaller than the lab-frame time by
a factor $\gamma^2$. Thus, the isotropic luminosity due to one 
electron, in the observer frame, is
given by
\begin{equation}
  \delta L_{iso} \approx \gamma^4 p_e \approx {2 q^2 \gamma^8 c\over 3 \rho^2}
    \sim (2\times 10^{-7}{\rm \: erg\,s}^{-1})\; \nu_9^{8/3} \r5^{2/3}.
    \label{dL_iso}
\end{equation}

Since we are considering radiation from a coherent patch where all electrons
are moving with almost exactly the same velocity ($\gamma$) and streaming 
along nearly parallel B-field lines, we can take the patch to be 
moving with a Lorentz factor $\gamma$, and electrons in the patch's comoving 
frame to be essentially at rest. Let us consider that the electron density in
the comoving frame is $n_e'$, which we express in terms of 
the Goldreich-Julian (GJ) density $\nGJ$ \citep{1969ApJ...157..869G}:
\begin{equation}
  n_e' = \xi \, \nGJ,  \quad \mbox{and in the lab frame} \quad
  n_e = \gamma n'_e \quad, 
    \label{nep}
\end{equation}
where
\begin{equation}
 \nGJ \approx {2B_0\over q c P} \left({R_{ns}\over R}\right)^3 \sim (10^{14} 
    {\rm \: cm}^{-3})\, P_{-1}^{-1} B_{0,14} \left({R_{ns}\over R}\right)^3.
    \label{nGJ}
\end{equation}
$R_{ns}$, $P$ and $B_0$ are the radius, rotation period and surface
B-field strength of a neutron star, and $R$ is the distance from the
coherent patch to the center of the star.
The spin down time for the neutron star due to dipole
radiation is 
\begin{equation}
    t_{sd} \sim {M_{ns} P^2 c^3\over \pi^2 B_0^2 R_{ns}^4}
      \sim (250\,{\rm yr})\, P_{-1}^2 B_{0,14}^{-2} R_{ns,6}^{-4}.
    \label{ns-spin-down}
\end{equation}
So if the object underlying FRBs is a neutron star with B-field
strength $\sim10^{14}$ G, then it can produce radio bursts for only a 
relatively short period of time of order a few hundred years. We note
that the surface B-field $B_0$ of the FRB progenitor star may not be
of dipole configuration; higher order multipole components give a
smaller spin down power and hence a longer spin down time.

As our fiducial parameter, we take
$\xi\sim10$ and use $n_e' = 10^{15}n_{e,15}'\rm\ cm^{-3}$ hereafter,
so the total number of electrons in the
patch that are radiating coherently is
\begin{equation}
    N_e = n_e' {V}'_{coh} = \eta n_e' (\gamma\lambda)^3 \approx
    6\times 10^{23} \eta\nu_9^{-2}\rho_5 n_{e,15}'.
    \label{Ntot}
\end{equation}

The isotropic equivalent luminosity, in the observer frame, due to
$N_{\rm patch}$ coherent patches (each containing $N_e$ electrons) can
be calculated using equations 
(\ref{dL_iso}) \& (\ref{Ntot})
\begin{equation}
    L_{iso} = N_{patch} N_e^2 \delta L_{iso} \approx {2 c^7 q^2
    \eta^2 \gamma^{14} (n_e')^2 N_{patch} \over 3 \nu^6 \rho^2}, 
    \label{Liso1}
\end{equation}
or 
\begin{equation}
    L_{iso} \sim (10^{41} {\rm\ erg\ s}^{-1}) N_{patch}
     \eta^2 \nu_9^{-4/3} \r5^{8/3} (n_{e,15}')^{2}.
    \label{Liso2}
\end{equation}
In arriving at the above equation we made use of equation (\ref{gam}). The 
observed FRB isotropic luminosity is $\sim 10^{43}$ erg s$^{-1}$, which can 
be easily accounted for if $\eta\sim 10$, $\rho_5 \sim 6$ or $n_{e,15}'\sim 10$ (or
smaller values if some combination of 
these variables are adjusted). The beaming corrected luminosity in the
lab frame is
\begin{equation}
   L_{lab} \approx {L_{iso} (\Omega_F/4\pi)\over (\gamma^2/2)} \sim
    {c^7 q^2 \eta \gamma^{10}  (n_{e}')^{2}N_{patch} \over 3
      \nu^6 \rho^2}, 
    \label{L_lab1}
\end{equation}
\begin{equation}
     L_{lab} \sim (4\times10^{34} {\rm\ erg\ s}^{-1}) N_{patch} \xi^2 \eta
       \nu_9^{-8/3} \r5^{4/3}  (n_{e,15}')^{2}.
    \label{L_lab2}
\end{equation}
The factor in the numerator on the RHS of equation (\ref{L_lab1}) is
the solid angle of the radiation beam from the patch divided by
$4\pi$, and the factor in the denominator is the ratio of the observer frame and lab frame times. 
 
The total kinetic energy of electrons in all radiating patches in the
lab frame is
\begin{equation}
    E_{ke} = \gamma m_e c^2 N_e N_{patch} \sim (1.5 \times 10^{19} {\rm\ erg}) 
   {\eta N_{patch} \r5^{4/3}  n_{e,15}' \over \nu_9^{5/3} }.
    \label{E_e_tot}
\end{equation}
Combining equations (\ref{L_lab2}) \& (\ref{E_e_tot}), we find the cooling
time of electrons in the lab frame
\begin{equation}
    t_{cool} \sim {E_{ke}\over L_{lab}} \sim (3\times10^{-16} {\rm\ s})
         \, \nu_9  (n_{e,15}')^{-1},
    \label{tcool}
\end{equation}
which is much shorter (by a factor $\sim10^{9}$) 
than the lab frame time duration ($\gamma^2\nu^{-1}$) over which the
patch is radiating. This requires that there is an electric field parallel
to the B-field that can accelerate electrons and sustain their
Lorentz factor at $\gamma$ for the time duration of $\sim \gamma^2 \nu^{-1}$. The required electric field in the lab frame is given by
 \begin{equation}
     q E_\parallel \times c t_{cool} \sim \gamma m_e c^2 \; {\rm or} \;
        E_\parallel \sim (5\times10^{9} {\rm\ esu}) {\r5^{1/3}
        n_{e,15}'\over \nu_9^{2/3}}.
     \label{E_parallel}
 \end{equation}
We can express the electric field in terms of the FRB luminosity using
equation (\ref{Liso2})
\begin{equation}
   E_\parallel \sim (7\times10^{10} {\rm\ esu}) 
     L_{iso,43}^{1/2} N_{patch}^{-1/2} \eta^{-1} \r5^{-1}.
    \label{E_parallel1}
\end{equation}
This field, parallel to the B-field, is of the same order as
the perpendicular electric field associated with the FRB radio emission 
(equation \ref{E_em}).  The radiation can be sustained for
a duration longer than $\nu^{-1}$ in the observer's frame provided that
the electric field is maintained by some process for longer than
$\gamma^2\nu^{-1}$ (lab frame) such as an ongoing reconnection of
B-field. In \S\ref{current-sheet}, we provide a
possible magnetic reconnection geometry in which such a parallel
electric field can be maintained. 

The electric current associated with the motion of electrons in the
volume $V_{coh}$ is $I\sim q \gamma n_e'\eta (\lambda\gamma)^2 c$. The
number density of particles with positive charge is nearly the same as
the electron number density in order that  
the plasma is charge neutral. And since these positively charged particles 
are accelerated by the electric field in opposite direction to the electrons, 
the total current in the region is $2I$. Thus, the B-field generated 
by the current inside the patch is
\begin{equation}
   B_{ind} \sim {4 \pi I\over c (\eta_x+\eta_y) \lambda\gamma} \sim 2\pi q
   \eta^{1/2}  n_{e}' \lambda\gamma^2,
    \label{B_ind}
\end{equation}
or
\begin{equation}
    B_{ind} \sim (7\times10^{10}{\rm\ G}) {\eta^{1/2}
     \r5^{2/3}  n_{e,15}'\over \nu_9^{1/3}} \sim (10^{12}{\rm\ G}) {
       L_{iso,43}^{1/2} \nu_9^{1/3} \over N_{patch}^{1/2} 
       \eta^{1/2}\r5^{2/3}}.
    \label{B_ind1}
\end{equation}

This induced B-field is in the direction perpendicular to the primary
B-field ($\vec B_0$), and the induced field lines are closed curves 
that lie in planes normal to $\vec B_0$. Thus the superposition of the
two fields -- $\vec B_{ind}$ \& $\vec B_0$ -- will point in different
directions at different locations in the source. Particles are stuck in the
lowest Landau state, as shown later, and therefore their motion is 
along B-field lines. In order for these particles to produce coherent 
radiation their velocity vectors should be nearly parallel. To be more 
precise, the angle between the velocity vectors for different particles 
should not be larger than $\gamma^{-1}$. This constraint provides a lower 
limit on the strength of the original B-field $B_0$ in the region:
\begin{equation}
   B_0 \gae \gamma B_{ind} \sim (5\times10^{13} {\rm\ G}) L_{iso,43}^{1/2} 
     N_{patch}^{-1/2} \eta^{-1/2}\nu_9^{2/3} \r5^{-1/3}.
   \label{B0-lim}
\end{equation}

It is unphysical that a large number of patches would turn on 
{\it simultaneously} on a timescale of $\nu^{-1}\sim1$ ns, so we expect 
that $N_{patch}\sim 1$. Moreover, $\eta^{1/2}$ is at most of order
unity, as been constrained in \S\ref{sec:general}
(eq. \ref{eq:16}). This means that FRBs are most likely produced near
the surface of a neutron star or the event horizon of a stellar-mass
black hole in a region where the B-field strength is $\gtrsim 10^{14}$ G.

% due to the following two arguments. First, the FRB beaming factor
% $\Omega_{\rm F}/4\pi \sim (4\eta\gamma^2)^{-1}$ decreases with increasing
% $\eta$. For example, if $\eta^{1/2}=5$, we have $\Omega_{\rm F}/4\pi
% \sim 10^{-5} \nu_9^{-2/3}\rho_5^{-2/3}$, which means only an extremely
% small fraction 
% of all FRBs are seen by the observer on a certain line of
% sight. Second, if the NS has rotational period $P$, the time it takes
% for the beaming angle
% $2\gamma^{-1}\mbox{min}(1, \eta^{-1/2})$ to sweep across the
% line of sight must be longer than the FRB duration $\delta t$,
% i.e.\footnote{The constraint $P\gtrsim 0.1\rm\ s$ gives an upper limit
%   on the rotational energy of the NS, $E_{\rm rot} \lesssim
%   2\times 10^{48}\rm\ erg$. } 
% \begin{equation}
%   \label{eq:17}
%   P \gtrsim (0.1\rm\ s)\delta t_{-3}
%   \nu_9^{1/3}\rho_5^{1/3}\mbox{max}(1, \eta^{1/2}).
% \end{equation}
% This means $\eta\gg 1$ would require the NS to have a long period, which 
% is difficult to reconcile with the requirement on $n_e$ (see eqs. 
% \ref{nGJ} \& \ref{Liso2}) and radio observations of magnetars 
% in our galaxy. Based on the above arguments, we conclude that 
% the primary B-field strength (in equation \ref{B0-lim}) is unlikely
% to be much less than $5\times10^{13}$ G.

\subsection{Radio wave propagation through neutron star magnetosphere \& 
    constraint on particle density $n_e$}
\label{ffabs}

In this sub-section we discuss effects on GHz waves as they propagate through
the magnetosphere of a neutron star. The main goal is to ascertain whether 
radio waves can propagate through this media without suffering excessive 
absorption, and to provide a constraint on the particle density in the
region where FRB radiation is produced. We show below that wave
absorption is small as long as the particle number density ($n_e$) is
smaller than 10$^{18}$ cm$^{-3}$, which is about $10^2$ times the GJ
density, i.e. $\xi \lae 10^2$ (note that $n_e' = 
\xi \nGJ$ and $n_e = \gamma n_e'$).

Transverse electromagnetic waves produced by the curvature radiation are
linearly polarized with the electric field vector ($\vec E_{EM}$) perpendicular
to both the local B-field ($\vec B_0$) and the wave-vector $\vec k$
as long as $\omega_c \gg \omega_e$, where
\begin{equation}
    \omega_e^2 = {4\pi q^2 n'_e\over m_e} \quad {\rm or}\quad \omega_e \approx
   (2\times10^{12}\,{\rm rad\,s}^{-1})(n_{e,15}')^{1/2},
\end{equation}
is the plasma frequency, and 
\begin{equation}
   \omega_c \equiv {q B_0\over m_e c} = (1.8\times10^{21}\,{\rm rad}\, 
    {\rm s}^{-1}) B_{0,14},
\end{equation}
is electron cyclotron frequency. 

The condition for the propagation of transverse EM waves, with $\vec
E_{EM}$  
perpendicular to $\vec B_0$, in a highly magnetized plasma, is:  $\omega > 
\omega_e^2/\omega_c$ \citep[e.g.][]{1986ApJ...302..120A}. Therefore,
although the plasma frequency in the FRB  
source region is much larger than FRB radio frequency, the GHz waves can
still propagate through the medium at a speed that is extremely close to $c$.

It can be shown that as radio waves propagate through the magnetosphere
with changing B-field direction, $\vec E_{EM}$ adiabatically changes
its direction so that it is perpendicular to both the local $\vec B_0$ \& 
$\vec k$, as long as $\omega_e \gg \omega$ \citep[see
][]{1979ApJ...229..348C}.

In other words, the polarization vector of photons is along $\vec k \times 
\vec B_0$ while moving through the region of the magnetosphere where 
$\omega_e \gg \omega$.

The column density of electrons in the source region of FRB model we 
have described above is $\sim n_e' \gamma^3\lambda\sim 10^{21}\
n_{e,15}$ cm$^{-2}$ (or 300 pc cm$^{-3}$). However, the contribution of the magnetosphere 
to the observed DM value of FRBs is negligible. This is due to the fact that 
the {\it effective plasma frequency} in the presence of a strong B-field 
is reduced by a factor of $(\omega_c/\omega)^{1/2} \sim 10^{5}$, and thus
the contribution to the DM from the FRB source region is $\ll 1$ pc
cm$^{-3}$.  

\subsubsection{Free-free absorption of GHz radiation}

The FRB waves propagating through the medium where they are produced
are subject to the free-free absorption process, which is calculated here.
The standard result for free-free absorption
\citep[e.g.][]{1979rpa..book.....R} cannot be used for the FRB system
because particle motion is restricted 
by the strong magnetic field and is essentially confined to one dimension.
A self-consistent derivation of free-free opacity for the FRB source is 
presented below.

Consider Coulomb interaction between an electron ($e^-$) and a proton 
(or positron) -- hereafter $p^+$ -- moving with Lorentz factors $\gamma_e$ 
and $\gamma_p$, respectively. These particles are streaming along strong 
magnetic field lines of strength $B_0$, and they remain in the lowest 
Landau state during the entire interaction. The angle between their 
momentum vectors is $\theta$ in the lab frame. We will use the method 
of virtual quanta to calculate Bremsstrahlung emission and absorption, 
which is carried out in the rest frame of the electron (prime frame), 
and then transformed back to the lab frame. The Lorentz factor of $p^+$ 
and the angle between the momentum vectors in the prime frame are given by
\begin{equation}
   \gamma_p' = \gamma_p\gamma_e (1 + \beta_p\beta_e\cos\theta), \quad
       \beta_p'\gamma_p'\sin\theta' = \beta_p\gamma_p\sin\theta,
  \label{gam-prime}
\end{equation}
where $\beta_p' = v_p'/c$.

The x-axis is taken to be along the electron-momentum, and the x-y
plane is the plane of $e^-$ \& p$^+$ momenta. The magnetic field in
the prime frame at the location of the electron is
\begin{equation}
    B_x' = B_0, \quad\quad  B'_y = B'_z = 0.
\end{equation}

The main idea behind the virtual quanta technique is that the
electromagnetic 
field due to $p^+$ in the prime frame is sharply peaked in a narrow region
close to the particle. This field is Fourier decomposed and each frequency 
component is treated as a virtual photon that gets scattered by 
the electron. 

The electric field components due to $p^+$ in the prime frame is calculated 
using the Li\'enard--Wiechert potential. We prefer to calculate the field 
in the $p^+$ rest frame and then Lorentz transform it to the prime frame followed 
by a rotation along the z-axis by an angle $\theta'$ so as to obtain field 
components along the $x'$ and $y'$ axes. The $x'$ component of the electric 
field at the location of the electron is given by 
\begin{equation}
   E'_{x'} = -{q\gamma_p'\over {r'_e}^3} \left[b'\sin\theta' + v_p' t' 
      \cos\theta'\right],
\end{equation}
where
\begin{equation}
   r'^2_e = b'^2 +  \gamma'^2_p v_p'^2 t'^2,
\end{equation}
and $b'$ is the minimum distance between $e^-$ and $p^+$ in the prime frame.

For strong $B_0$ field associated with the FRB model, the y-component
of the electric field and the magnetic field associated with $p^+$ are too
weak and of too small a frequency to excite the electron to a higher Landau
level and therefore these fields are ignored for the Bremsstrahlung
calculation. The electron motion along the x'-axis is 
described by
\begin{equation}
    \vec \beta'_e = -{q E'_{x'} \over m_e c} \hat x'.
\end{equation}

The electric field in the radiation zone due to the acceleration of 
the electron is
\begin{equation}
    \vec{E}'_{rad} = {q\, \hat n'\times \left[ (\hat n' -
        \vec{\beta}_e')\times 
       \dot{\vec\beta}'_e\right] \over c r' (1 - \hat n'\cdot \vec\beta'_e)^3}
     \approx {q \over c r'}\left[ \hat n' \cos\psi' - \hat x'\right] 
     |\dot{\vec{\beta}}'_e|,
\end{equation}
where 
\begin{equation}
   \cos\psi' = \hat n'\cdot \hat x'.
\end{equation}

The Poynting flux due to the electron acceleration is given by
\begin{equation}
   \vec S' = {c\over 4\pi} \vec E'_{rad} \times \vec B'_{rad} = {|\vec E'_{rad}
     |^2 c \,\hat n' \over 4 \pi}.
    \label{s-prime}
\end{equation}
The energy release per unit frequency and per unit solid angle in a
$e^-$--$p^+$ scattering is
\begin{equation}
   {d\epsilon'\over d\omega' d\Omega'} = {q^2 \sin^2\psi'\over 2\pi c}
    |\dot{\hat{\beta}}'_e(\omega')|^2, 
\end{equation}
where
\begin{equation}
   \dot{\hat{\beta}}'_e(\omega') = {q^2 \gamma'_p\,\hat x'\over (2\pi)^{1/2}
    m_e c} \int_{-\infty}^\infty dt'\, e^{i\omega' t'} {( b' \sin\theta' + 
     v'_p t' \cos\theta')\over (b'^2 + \gamma'^2_p v'^2_p t'^2)^{3/2}}. 
\end{equation}

This integral can be carried out using the modified Bessel functions
\begin{align}
    K_1(a) & = {a\over 2} \int_{-\infty}^\infty dz\, {e^{i z}\over
        (a^2 + z^2)^{3/2}},  \\
    K_0(a) & = {1\over 2 i}  \int_{-\infty}^\infty dz\,{z 
    e^{iz}\over (a^2 + z^2)^{3/2}}, 
\end{align}
and we obtain 
\begin{equation}
\begin{split}
   \dot\beta'_e(\omega') = & {2 q^2 \omega'\over (2\pi)^{1/2} m_e c^3 (\beta'_p
      \gamma'_p)^2)} \Bigg[ \gamma'_p\sin\theta' K_1\left({b'\omega'\over 
      v'_p\gamma'_p}\right) \\
    & \quad\quad\quad\quad\quad\quad\quad\quad + \; i \cos\theta' 
       K_0\left({b'\omega'\over v'_p\gamma'_p}\right)\Bigg],
\end{split}
    \label{acc-prime}
\end{equation}
where 
\begin{equation}
   \omega' = \omega \gamma_e (1 - \beta_e \cos\psi)\equiv {\cal
     D}^{-1} \omega,  \quad \sin\psi' = {\cal D} \sin\psi.
\end{equation}

Making use of equations (\ref{s-prime}) \& (\ref{acc-prime}), we obtain the 
energy release per unit frequency and per solid angle in one $e^-$--$p^+$ 
encounter 
\begin{equation}
  {d^2\epsilon\over d\omega d\Omega} = {\cal D}^2 {d^2\epsilon'\over 
    d\omega' d\Omega'}, 
\end{equation}
or
\begin{equation}
\begin{split}
    {d^2\epsilon\over d\omega d\Omega} = & {q^6 \omega^2 \sin^2\psi
     \ {\cal D}^2 \over \pi^2 m_e^2 c^3 (v'_p\gamma'_p)^4 }\; \times \\
      & \quad \bigg[ {\gamma'_p}^2\sin^2\theta' K_1^2\left({b'\omega'\over
      v'_p\gamma'_p}\right) + \cos^2\theta' K_0^2\left({b'\omega'\over
        v'_p\gamma'_p}\right)\bigg]. 
\end{split}
\end{equation}

For a plasma with $e^-$ and $p^+$ densities of $n_e$ \& $n_p$ the 
Bremsstrahlung emissivity is given by
\begin{equation}
   j^{ff}(\omega,\Omega) = n_e n_p 2\pi \int db\, b\, v_p 
      {d^2\epsilon\over d\omega d\Omega}. 
\end{equation}
The emission is beamed along the electron momentum vector within an angle
of $\gamma_e^{-1}$, and the spectrum peaks at $\omega \sim \gamma_e\gamma'_p
(c/b') \sim \gamma_p \gamma^2_e c n_e^{1/3}$ which is a factor $\sim 10^{10}$
larger than the GHz frequency of FRBs we are interested in. Therefore, we will
consider the low frequency limit where
\begin{equation}
   K_0(a)\approx -(\ln[a/2] + \xi^\prime),\quad\quad
        K_1(a) \approx 1/a,
\end{equation}
$\xi^\prime=0.5772$ is the Euler-Mascheroni constant. It can be shown that for
FRBs, the term containing $K_0$ in the above expression for $ j^{ff}$ is
much smaller than the $K_1$ term. With these approximations, we finally 
arrive at the following simplified expression for Bremsstrahlung emissivity
\begin{equation}
    j^{ff}(\omega,\Omega)\approx {8 q^6 n_e n_p\sin^2\theta\over \pi m_e^2 c^4} 
    {(\gamma_e\psi)^2\over \left[1 + (\gamma_e\psi)^2\right]^4}
       \ln\left[{b_{max}\over b_{min}}\right],
\end{equation}
where $\psi$ is the angle between electron momentum vector and the photon,
and $\theta$ is the angle between $e^-$ \& $p^+$ momentum vectors; 
$b_{max}/b_{min}\sim m_e c^2 \gamma_e^3/(\omega \hbar)$. The emissivity
decreases very rapidly with angle for $\psi > \gamma_e^{-1}$.

Finally, we can obtain the expression for free-free absorption coefficient,
including correction for the stimulated emission, in Rayleigh-Jeans limit 
using the Kirchhoff’s law:
\begin{equation}
  \kappa^{ff}(\nu,\psi) = j^{ff}_\nu/B_\nu(T)\approx {\pi j^{ff}_\omega c^2\over
     k_B T \nu^2},
\end{equation}
or
\begin{equation}
    \kappa^{ff}(\nu,\psi) \sim (5{\times}10^{-15} {\rm\; cgs\ units)\ }{(\gamma_e\theta)^2 
    (\gamma_e\psi)^2 n_e^2 \over \gamma_e^{3} \nu^{2} \left[1 + 
    (\gamma_e\psi)^2\right]^4},
    \label{kff1}
\end{equation}
where the electron temperature $k_B T\sim m_e c^2 \gamma_e$ along
  the magnetic field.

Inside the source region where the photons are moving almost parallel
to electrons that produced them --- $\psi\gamma_e\sim 1 \sim \theta\gamma_e$ ---
 the optical depth corresponding to the source length $\ell$ along the line 
of sight is
\begin{equation}
  \tau^{ff} \approx \kappa^{ff}_\nu \ell /(2\gamma_e^2) \sim 10^{-2}
  \, \ell_5\,
     n_{e,17}^2 \nu_9^{-2},
   \label{tau-ff}
\end{equation}
where we took $\gamma_e\sim 30$ as per equation (\ref{gam}). The optical 
depth due to particles moving in the direction
opposite to the electrons is much smaller than that given by equation 
(\ref{tau-ff}) for a electron-positron plasma\footnote{This
is due to the fact that free-free radiation produced by
positrons is narrowly beamed along their momentum vector, and very little
radiation comes out in the opposite direction. Absorption of photons
moving head-on toward positrons, which is related to the inverse of
the free-free emission, has therefore, much smaller cross-section than
absorption of photons moving in the same direction as positrons. The lower
cross-section more than compensates for the larger number of positrons
encountered by photons while traveling the distance $w$.}.
For a electron-proton plasma, the absorption of GHz photons by protons
is even smaller and can be ignored.

Equation (\ref{tau-ff}) suggests that the particle density within the 
FRB source region should not be much larger than $\sim 10^{18}$ cm$^{-3}$
to avoid free-free absorption of photons.
However, the allowed density is sufficient for producing the observed 
typical FRB luminosity of $L_{iso} \sim 10^{43}$ erg s$^{-1}$ for a
coherent source of transverse size $\sim \gamma\lambda$, i.e. $\eta\sim 1$
 (see eq. \ref{Liso2}).

Once particles leave the acceleration zone the waves might suffer 
significant absorption if the outside temperature is much lower than 
$m_e c^2 \gamma_e/k_B$. This is, however, unlikely. The cooling time 
for electrons increases dramatically outside the {\it coherence zone}, 
and becomes of order $m_e c^2\gamma/p_e \sim (10^8$ s) $\nu_9^{-1} \r5$ 
(where $p_e$ is taken from equation \ref{pe}). In this case, the electron 
temperature stays relativistic and the free-free absorption 
at $\sim$ GHz is small as long as the {\it coherence zone} is contained 
inside a somewhat larger acceleration zone. Moreover, as photons travel
further out in the magnetosphere, the angle between their momentum vector and
the local magnetic field direction ($\psi$) increases, and 
$\kappa^{ff}$ decreases rapidly for $\psi \gae \gamma_e^{-1}$ 
(as $\psi^{-6}$ -- see eq. \ref{kff1}) which keeps the free-free 
absorption small.

The free-free absorption increases with decreasing frequency as $\nu^{-2}$, 
and therefore waves with $\nu \lae 500$ MHz are likely to be absorbed in the source
region. This might explain the lack of detection of FRBs below GHz frequencies
in a few cases.

The bottom line is that the GHz photons can escape the magnetosphere of
a magnetar without suffering much free-free absorption provided that the 
plasma density in the source region $n_e\lae 10^{18}$ cm$^{-3}$. This
upper limit to the density is larger than the GJ density 
by a factor $\sim 10^2$.
Moreover, the density required for our model to produce the observed
luminosity of a typical FRB of $L_{iso}\sim 10^{43}$ erg s$^{-1}$ is
$n_e\sim 10^{17}$ cm$^{-3}$, and the transverse size of the source is 
$\ell_t\sim \gamma\lambda$. 
The contribution to the DM of a FRB from the
magnetar source region is negligible due to the modification of the
EM wave dispersion relation in a highly magnetized plasma
($\omega_e/\omega_c\ll 1$).

\subsection{Energy dissipation in current sheet to power FRBs}
\label{current-sheet}

What we have established thus far, using general arguments, is that the
B-field in the region where FRB GHz radiation is produced is 
$\gtrsim 10^{14}$ G, and the electric field component parallel to the
B-field is $E_\parallel\sim 10^{11}$ esu. This electric field needs to be 
sustained for at least 1 $\mu$s ($\gamma^2\nu^{-1}$). In this sub-section 
we consider one possible way that such an electric field could arise:
when inclined B-field lines in the magnetosphere of 
a neutron star are forced to come together and reconfigure. In
this process of magnetic reconnection, a fraction of the B-field
energy is dissipated and charged particles are accelerated by $E_\parallel$ 
to relativistic speeds and produce coherent curvature radiation. 

The calculations presented in this sub-section are less robust than other 
parts of the paper because the physics of reconnection and particle 
acceleration are poorly understood especially when the magnetization
$\sigma = B_0^2/(4\pi n_p' m_p c^2)$ is large ($\sigma\gtrsim
10^{15}$ in the  FRB source region). Fortunately, the uncertainties
associated with the 
calculation  of reconnection and $E_\parallel$ do not affect other
parts of this paper apart from some of the predictions of our model
described in \S\ref{predict}. 

\begin{figure}
  \centering
\includegraphics[width = 0.45 \textwidth,
  height=0.23\textheight]{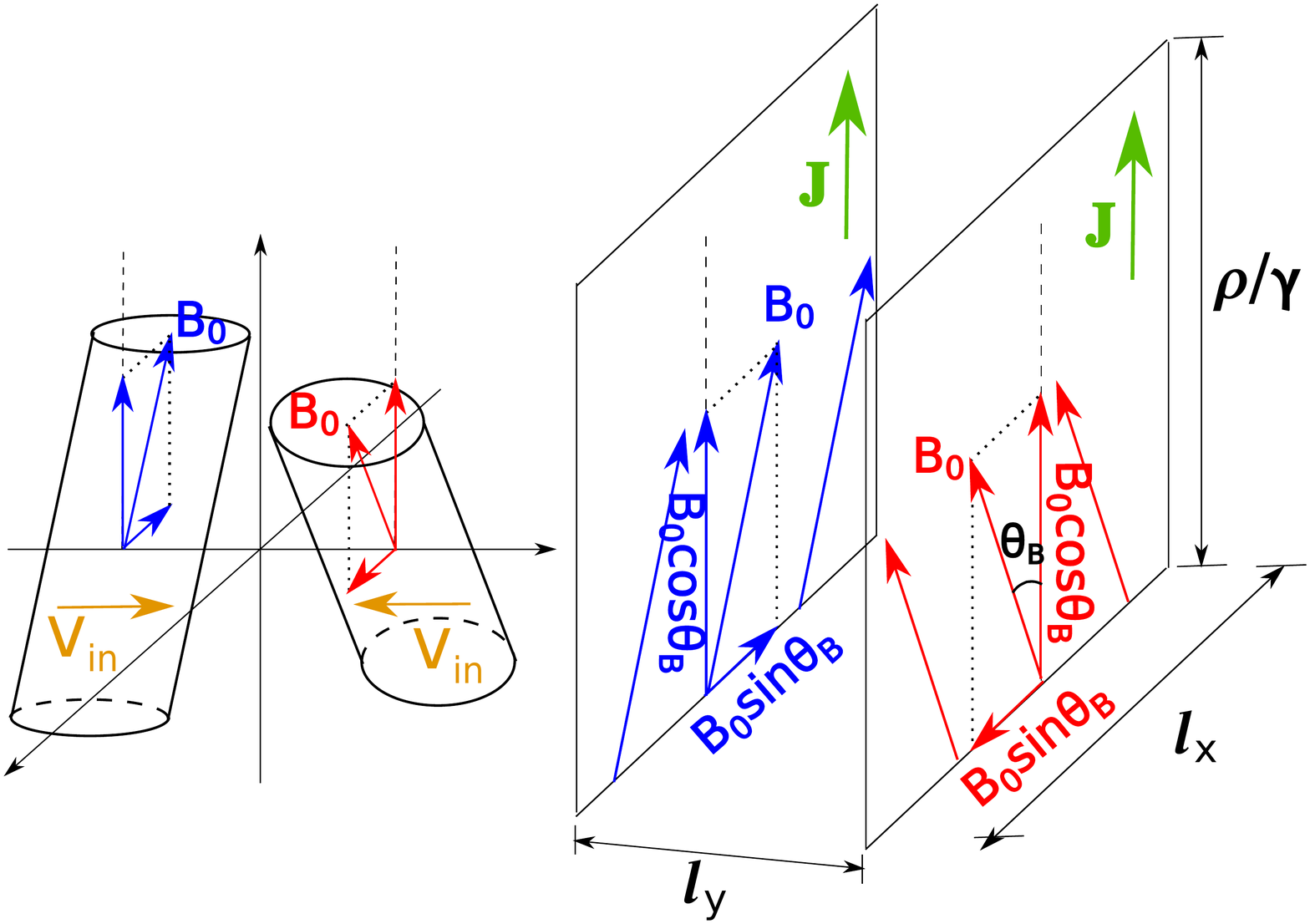}
\caption{Geometry of the current sheet during magnetic
  reconnection. Plasma with inclined magnetic fields flow in at
  speed $v_{\rm in} = \beta_{\rm in}c$ from opposite sides into the
current sheet. The
  electric field due to reconnection is $E_{\parallel} \sim
  B_0\sin\theta_{\rm B} \beta_{\rm 
  in}$ and is parallel to $B_0\cos \theta_{\rm B}$ (which is roughly
parallel to the line of sight). The dimension of the current sheet along
the line of sight is $\gtrsim \rho/\gamma$ and the transverse dimension
is $l_{\rm x}\gtrsim \eta^{1/2}\gamma\lambda$. The thickness
of the current sheet is $l_{\rm y}\gtrsim \eta^{1/2}\gamma\lambda$. In
this way, the transverse area of the coherent volume is $A_{\rm
  coh}=\eta\gamma^2\lambda^2$.
}\label{fig-current-sh}
\end{figure}

Let us consider the B-field directions at the opposite sides 
of the current sheet to be inclined at an angle $2\theta_B$, and B-field
strength to be $B_0$; we consider $\theta_B \ll 1$ rad. The B-fields on the
two sides of the sheet can be decomposed as parallel 
and anti-parallel components (see Fig. \ref{fig-current-sh}). Only the
anti-parallel components of the B-field, which have strengths of
$B_0 \sin\theta_B$, are dissipated inside the current sheet, whereas 
the parallel components with strength $B_0\cos\theta_B$ (or the guide
field) remain intact. Let us take the plasma speed flowing into the
current sheet to be $\beta_{\rm in} c$. The dimension of the
 current sheet along the line of sight (or along the guide field) is
  $\gtrsim \rho/\gamma$ and in the transverse 
direction is $l_{\rm x}\gtrsim \eta^{1/2}\gamma\lambda$. The thickness
of the current sheet is $l_{\rm y}\gtrsim \eta^{1/2}\gamma\lambda$. Thus,
the transverse area of the coherent volume is $A_{\rm coh}=
\eta\gamma^2\lambda^2$.

The dissipation rate of magnetic energy is 
\begin{equation}
  \label{eq:100}
  P_{\rm cs} \simeq \frac{(B_0\sin\theta_{\rm B})^2}{4\pi}\frac{\rho}{\gamma}
  \eta^{1/2}\gamma\lambda\beta_{\rm in} c = \frac{E_{\parallel}^2}{4\pi
    \beta_{\rm in}}\eta\rho\lambda c.
\end{equation}
From the cooling time argument, we require $E_{\parallel}\simeq
(L_{\rm iso}/c)^{1/2}\eta^{-1}\rho^{-1}$ and hence
\begin{equation}
  \label{eq:101}
    P_{\rm cs}\simeq \frac{L_{\rm iso}}{4\pi
    \beta_{\rm in}}\frac{\eta^{1/2}\gamma \lambda}{\eta^2 \gamma\rho}
  \simeq \frac{L_{\rm 
      lab}\gamma}{\beta_{\rm in}\eta^{1/2}}
\end{equation}
where we have used $L_{\rm lab} = L_{\rm iso}/(2\eta\gamma^4)$ and
$\lambda = 2\pi \rho/\gamma^3$. The radiation efficiency is
\begin{equation}
  \label{eq:102}
  f_r = \frac{L_{\rm lab}}{P_{\rm cs}}\simeq \frac{\beta_{\rm
      in}\eta^{1/2}}{\gamma}. 
\end{equation}
The inflow velocity is also given by $E_{\parallel} = B_0\sin \theta_{\rm
  B}\beta_{\rm in}\simeq (L_{\rm iso}/c)^{1/2}\eta^{-1}\rho^{-1}$, i.e.
\begin{equation}
  \label{eq:103}
  \beta_{\rm in}\sin \theta_B \simeq 2\times 10^{-3} B_{0,14}^{-1}
  L_{\rm iso, 43}^{1/2} \eta^{-1}\rho_5^{-1}.
\end{equation}

The angle $\theta_B$ can be roughly estimated from the condition that
$|\vec\nabla\times\vec B| < 4\pi J_{max}/c$; where $J_{max} \sim q n_e c$.
Since $|\vec\nabla\times\vec B|\sim B_0 \sin\theta_B/l_x$, we find 
that $\theta_B \lae 4 \pi q n_e l_x/B_0 \sim 10^{-2} n_{e,17}\eta^{1/2} 
B_{0,14}^{-1}$. The system becomes charge starved for larger angles,
and that would trigger rapid dissipation of magnetic fields. For
$\theta_{\rm B}\sim 10^{-2}$, we have $\beta_{\rm in}\sim 0.2$ and 
hence $f_r \simeq 7\times 10^{-3}\eta^{1/2}\rho_5^{-1/3}\nu_9^{-1/3}$.
There is considerable flexibility (and uncertainty) in values the parameters 
$\theta_{\rm B}$, $B_0$, $\eta$ and $\rho$ can take, and hence a wide
range of efficiencies from $10^{-4}$ to $10^{-1}$ are allowed in the 
magnetic reconnection model.

The intrinsic durations of most FRBs only have upper limits $\sim
1\rm\ ms$, because they were not resolved after deconvolution of the
scattering broadening and DM smearing resulting from the finite 
channel width of radio telescopes. However, the repeating
events from FRB 121102 showed various observed durations (2.8-8.7 ms)
with no 
clear evidence of scattering broadening
\citep{2016Natur.531..202S}. Also, FRB 
121002 showed resolved double-peaked profile with separation of about
$2.4\rm\ ms$ between the peaks \citep{2016MNRAS.tmpL..49C}. If we
assume that these relatively long durations are intrinsic, the process
that drives the magnetic reconnection may be relatively ``slow'',
i.e. operating on timescales much longer than the light-crossing time
of the NS. The large scale steady state structure of the
force-free NS magnetosphere (e.g. dipole or
  multipole with twists) can be solved
  \citep[e.g.][]{2016MNRAS.462.1894A}. However, the B-field
  configuration very close to the surface of a NS (most 
likely a magnetar for FRBs) is still poorly 
understood \citep[e.g.][]{2008A&ARv..15..225M}, so the mechanism for
forced 
reconnection of magnetic fields is uncertain. One possible scenario 
is that magnetic flux emerges from below the NS surface due to buoyancy
\citep[e.g.][]{1995ApJ...440L..77M,  
  2012MNRAS.425.2487V} and then the emergent B-field reconnects with
pre-existing B-field in the magnetosphere at some inclination. Another
possibility is the slow movement of the NS crust where the field lines
 are anchored. For example,
in the first scenario, flux emergence from under the NS surface occurs
on a timescale 
\begin{equation}
  \label{eq:11}
  t_{\rm em} \sim \frac{H}{v_{\rm A}} \sim (3{\rm\ ms}) H_4
  B_{0,14}^{-1} \rho_{0,14}^{1/2},
\end{equation}
where $H = 10^4 H_4\rm\ cm$ is the depth from which
the flux emerges, $\rho_0 = 10^{14}\rho_{0,14}\rm\ g\ cm^{-3}$ is
mass density of the surface layer, $v_{\rm A} =
B_0/(4\pi\rho_0)^{1/2}$ is the Alfv{\'e}n speed.

\subsection{Robustness of coherent curvature radiation}

In \S\ref{sec-particle}, we calculate the radiation field from a
single particle moving along a fixed B-field line, and then we
add up the contribution from particles with different Lorentz factors
and moving along magnetic field lines pointing in different directions. 
Taking $\rho = 10^5\rm\ cm$ as an example, we show that the curvature
radiation from different particles within a patch adds up coherently
provided that (i) dispersion of the Lorentz factor of particles in the 
region is within a factor a few of the mean Lorentz factor
($\gamma$); (ii) the  
orientations of magnetic 
field lines at different points in the region are within an angle 
$\gamma^{-1}$ of each other; (iii) the spectrum of the emergent coherent 
radiation can have strong fluctuations on $\sim$ MHz scale; (iv) the
time  
delay between radiation from opposite ends of the patch should be no 
larger than $\nu^{-1}$.

In \S3.3.2 \& 3.3.3, we discuss a number of
physical effects that could perturb particle trajectories, generate or
inhibit the generation of coherent bunches.

\subsubsection{Condition on particle velocity and B-field orientation
   distributions for coherent curvature radiation}
\label{sec-particle}

\begin{figure}
  \centering
\includegraphics[width = 0.45 \textwidth,
  height=0.22\textheight]{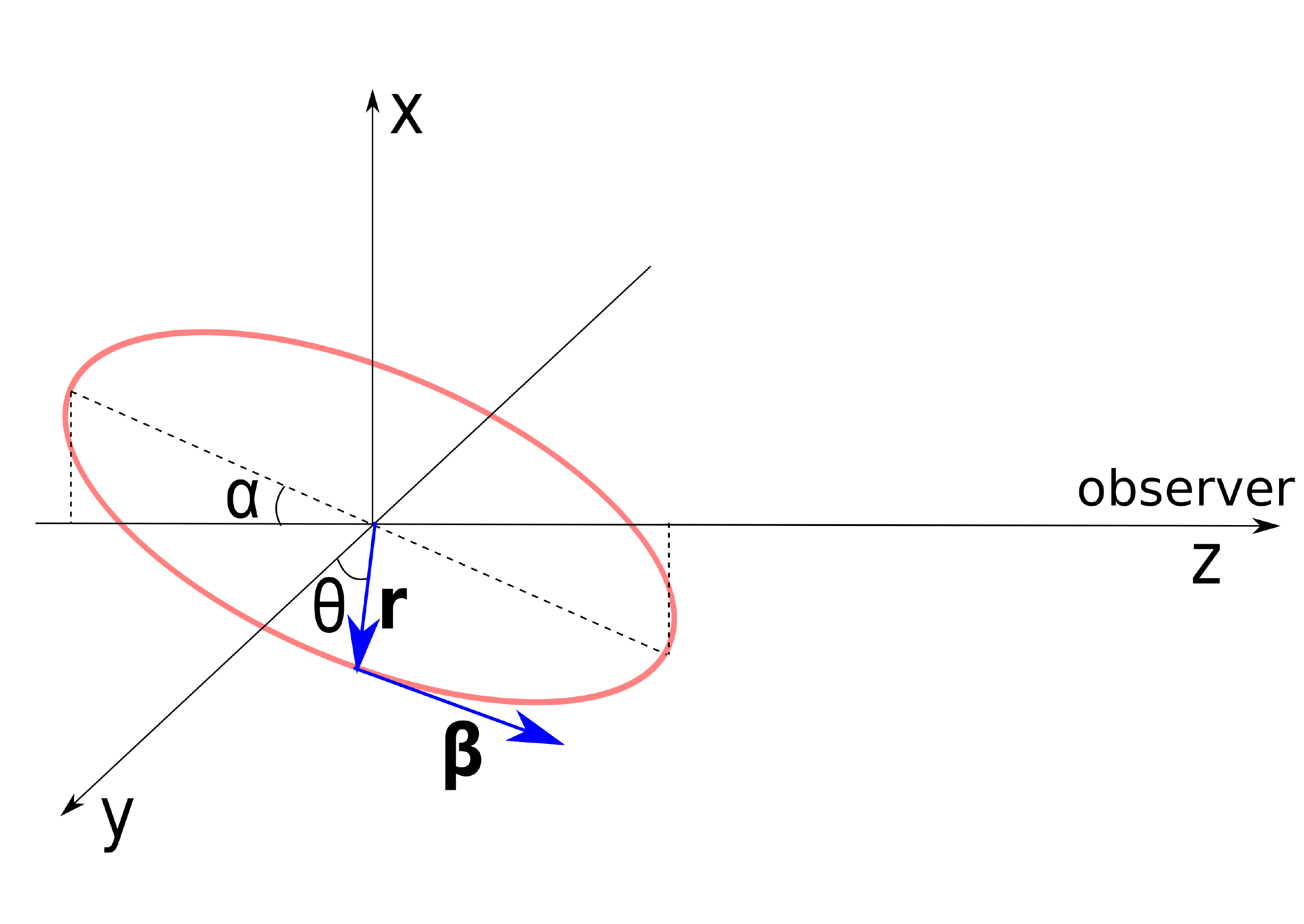}
\caption{Geometry of a single radiating particle in a circular
  orbit with radius $\rho$. The observer is at $d\ \hat{z}$ on the z
  axis (line of sight), $d$ being the distance. The orbital
  plane is tilted with respect to the z axis by an angle
  $\alpha$. The position of the particle within the orbital plane is
  parametrized by the azimuthal angle $\theta$, with $\theta=0$
  when the particle is on the y-axis. The position vector of the
  particle is $\vec{r}$ and the velocity vector (divided by $c$) is
  $\vec{\beta}$. The retarded time $t_{\rm r} = 0$ when the particle
  is on the x axis. The angular frequency is $\omega = \beta c/\rho$,
  so $\theta(t_{r}) = \omega t_{\rm r}$.
}\label{fig:geometry}
\end{figure}

% The main point of this sub-section is to show that the coherent curvature
% radiation does not require the distribution of particle Lorentz factor 
% along B-field lines to be narrow (a broad distribution of width
% $\sim\gamma$ works fine). However, the B-field lines in the source 
% region must be nearly parallel to each other, within an angle 
% of $\gamma^{-1}$, in order for radiation from different particles in 
% the region to add coherently.

Consider a single particle with charge $q$ on a
circular orbit of radius $\rho$ (the curvature radius of the B-field
line) inclined 
at an angle $\alpha$ with respect to the line of sight or $\hat{z}$
axis, as shown in Fig. (\ref{fig:geometry}). The angular frequency is
$\omega = \beta c/\rho$, where $\beta = |\vec{\beta}|$ is
the particle's speed divided by $c$ and we assume $\beta$ to be
constant with time in order to focus on the curvature radiation. The
azimuthal angle at any retarded time $t_{\rm r}$ is given by
$\theta (t_{\rm r}) = \beta  ct_{\rm r}/\rho$ (we have taken 
$t_{\rm r} = 0$ for $\theta = 0$). The electric field at
the observer's location $d\ \hat{z}$ due to the 
particle's acceleration is \citep{1979rpa..book.....R}
\begin{equation}
  \label{eq:6}
  \begin{split}
      \vec{E} & = \frac{q}{c\kappa^3 R} \vec{n}\times [(\vec{n} -
      \vec{\beta})\times \dot{\vec{\beta}}], \\
      &= \frac{q\beta\omega}{cd} \left[\frac{-\sin\theta \sin\alpha\
  \hat{x} + (\cos\theta - \beta\cos\alpha)\ \hat{y}}{(1 -
  \beta\cos\theta\cos \alpha)^3}\right]
  \end{split}
\end{equation}
where we have used $\kappa \equiv 1 - \vec{n}\cdot\vec{\beta}$, 
$\vec{n}\equiv \vec{R}/R$, $\vec{R}$ being the vector from the
particle's position to the observer, and we have neglected high order
terms $O(\rho/d)$.

\begin{figure}
  \centering
\includegraphics[width = 0.45 \textwidth,
  height=0.22\textheight]{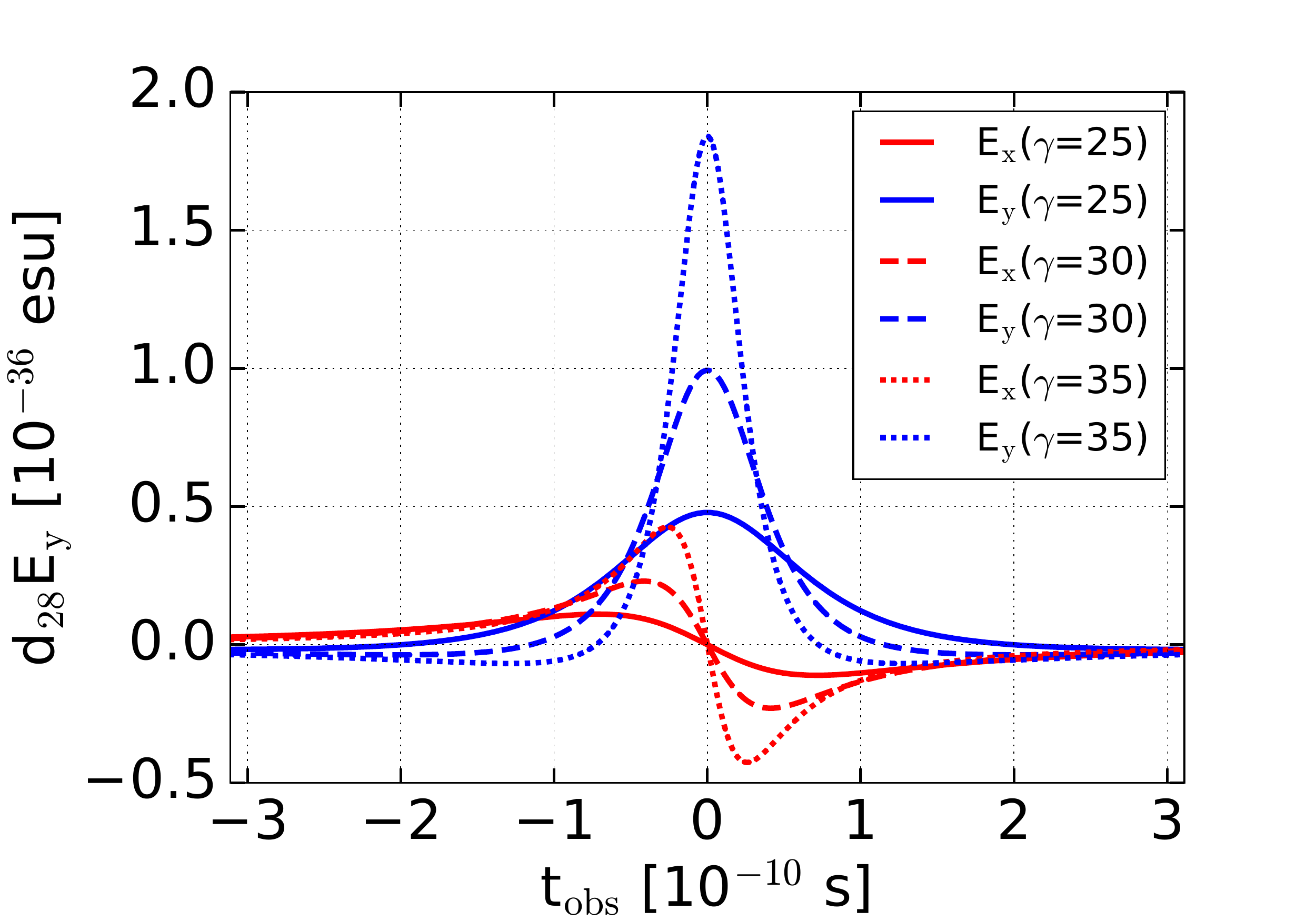}
\caption{The electric field from a single electron at a distance of
  $10^{28}d_{28}\rm\ cm$ according to equation (\ref{eq:6}). We consider
  three different Lorentz factors $\gamma = 25$ (solid
  lines), 30 (dashed lines), 35 (dotted lines); red lines are for
  $E_{\rm x}$ and blue lines are for $E_{\rm y}$ (see
  Fig. \ref{fig:geometry} for the definition of x and y axes). The 
  curvature radius is $\rho = 10^5\rm\ cm$ and the angle between the
  orbital plane and the line of sight is $\alpha = 0.5\gamma^{-1}$
  radian. If there are $N_{\rm e}$ electrons on an identical
  trajectory, the total electric field is $N_{\rm e}E$ and the
  isotropic luminosity is $L_{\rm iso} = d^2N_{\rm e}^2E^2c$. For $N_{\rm e} =
  10^{24}$, $E = 10^{-36}\rm\ esu$ at $d = 10^{28}\rm\ cm$ means
  $L_{\rm iso} = 3\times10^{42}\rm\ erg\ s^{-1}$.
}\label{fig:E_single}
\end{figure}

We show in Fig. (\ref{fig:E_single}) the x and y components of the electric
field in the far field due to a single particle. Note that $E_{\rm x}$
is anti-symmetric with $\theta$ and $\alpha$ while $E_{\rm 
  y}$ is symmetric. If the particle flow is continuous on a timescale
of $\gtrsim\nu^{-1}$, $E_{\rm x}$ from particles with positive and negative
$\theta$ will cancel out, but $E_{\rm y}$ will survive.
In the following, we only consider the $\hat{y}$ component of
$\vec{E}$, the magnitude of which is
\begin{equation}
  \label{eq:13}
  \begin{split}
      E_{\rm y}(t_{\rm obs}) &= \frac{q\beta\omega}{cd}
  \left[\frac{\cos\theta(t_{\rm r}) -
    \beta\cos\alpha }{(1 -
  \beta\cos\theta(t_{\rm r})\cos \alpha)^3}\right] \\
 &\simeq \frac{q\gamma^4}{\rho d} \mbox{, if both $\alpha$ and $\theta
   \ll \gamma^{-1}$},
  \end{split}
\end{equation}
where time $t_{\rm obs}$ in the observer frame is given by
\begin{equation}
  \label{eq:14}
  t_{\rm obs} = \int_{0}^{t_{\rm r}} (1 - \vec{n}\cdot
  \vec{\beta})\mathrm{d} t_{\rm r}
\approx \int_{0}^{t_{\rm r}} (1 - \beta \cos\alpha \cos\theta)
  \mathrm{d} t_{\rm r}.
\end{equation}
The emission is strong only when $\alpha \gamma \lesssim 1$ and the
duration of the emission from a single particle is $\Delta t_{\rm
  obs}\sim \rho /(\gamma^3c)$.

\begin{figure}
  \centering
\includegraphics[width = 0.45 \textwidth,
  height=0.35\textheight]{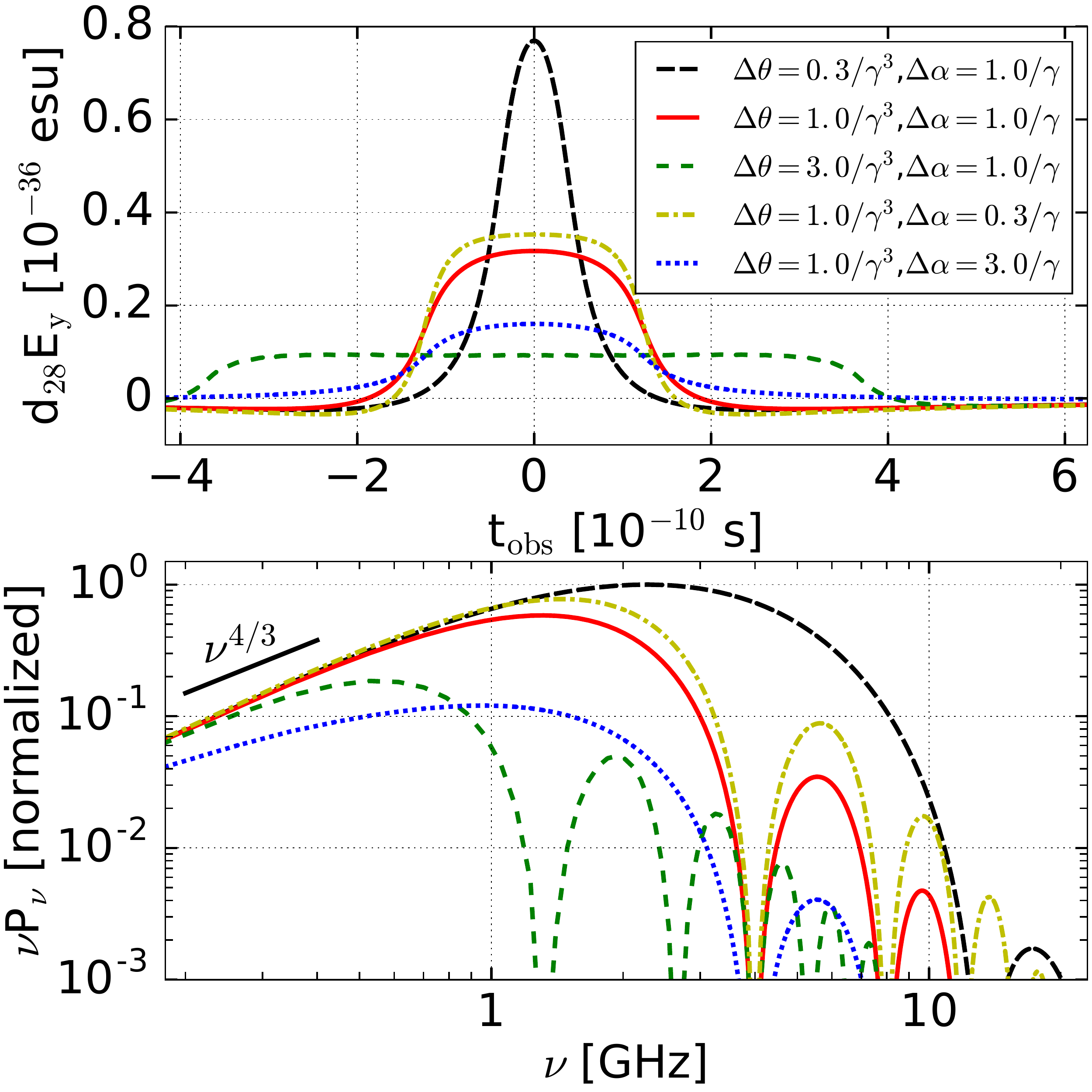}
\caption{{\it Upper Panel:} The electric field from a bunch of
  electrons with flat Lorentz factor distribution $\partial N/\partial
  \gamma = \rm const$ ($20<\gamma<40$), flat orbital plane angle
  distribution $\partial N/\partial \alpha = \rm const$ ($-\Delta \alpha <
  \alpha < \Delta \alpha$), and flat azimuthal angle delay distribution
  $\partial N/\partial \theta = \rm const$ ($-\Delta \theta< \theta
  <\Delta\theta$). The angle limits $\Delta \alpha$ and 
  $\Delta \theta$ are expressed as functions of the mean Lorentz
  factor $\gamma=30$ as shown in the legend. The total electric field
  is normalized to $N_{\rm e} = 1$ (to get the actual electric 
  field, one needs to multiply with the total number of electrons in the
  coherent patch). The source is at a distance of $10^{28}d_{28}\rm\
  cm$. Since $E_{\rm x}$ is 
  anti-symmetric with $\alpha$ and $\theta$ while $E_{\rm y}$ is
  symmetric, only $E_{y}$ will contribute to observations. The pulse
  width increases with $\Delta \theta$. The shape is 
  not strongly affected by $\Delta \alpha$, unless $\Delta \gtrsim
  1/\gamma$ (when the peak electric field is much lower because
  the emission from a large fraction of electrons is beamed away from
  the observer). {\it Lower Panel:} The frequency spectrum normalized to
  the peak value of the ($\Delta\theta = 0.3/\gamma^3, \Delta\alpha =
  1.0/\gamma$) case. The spectrum in the lower frequency end is
  $P_{\nu}\propto \nu^{1/3}$ and in the higher frequency end cuts off
  exponentially. For cases where $\Delta \theta \geq1.0/\gamma^3$,
  the pulse is like a step function; this is why the spectrum has
  regularly spaced discrete bumps \citep[which resemble the discrete
  band structure seen in Crab pulses][]{2007ApJ...670..693H}.
}\label{fig:coh_vol}
\end{figure}

In Fig. (\ref{fig:coh_vol}), we plot the $E_{\rm y}$ from a coherent
volume of electrons with flat distributions of Lorentz factors in the
interval $20<\gamma<40$, orbital angles between $-\Delta \alpha
< \alpha < \Delta \alpha$, and azimuthal angle in the interval $-\Delta
\theta < \theta_0 < \Delta \theta$. We consider three different
$\Delta \alpha$ of $0.3/\gamma$, $1.0/\gamma$ and $3.0/\gamma$
radian; and three different $\Delta\theta$ of $0.3/\gamma^3$,
$1.0/\gamma^3$, and $3.0/\gamma^3$; where $\gamma = 30$ is the mean
Lorentz factor. The electric field is normalized to $N_{\rm e} = 1$. To
get the actual electric field, one needs to multiply with the total
number of electrons in the coherent patch. The general conclusion is
that coherence is destroyed and the resulting electric field $E_{\rm
  y}$ is much weaker when the separation of particles in a patch
significantly exceeds $\lambda$ (i.e. $\Delta \theta\gg\gamma^{-3}$) or
when the direction of particle motion has a spread much larger than
$1/\gamma$ (i.e. $\Delta \alpha\gg \gamma^{-1}$). Particle clumps
with longitudinal size $\sim \lambda$ may be produced in various
plasma instabilities, such as the two-stream instability (see \S3.3.2).

We note that coherence at frequencies $\sim 1\rm\ GHz$ is
unaffected if particles have a broad Lorentz factor
distribution. The peak frequency of curvature radiation 
depends strongly on the particle Lorentz factor $\nu\propto 
\gamma^3$, so particles with Lorentz factors larger (or smaller) than
30 by a factor of more than 2 will radiate most of their energy at
frequencies much above (or below) $\sim 1\rm\ GHz$. During the
reconnection, particles with a wide range of Lorentz factors may be 
produced, but that would not affect the observed luminosity of $\sim 1$ Jy
at $\sim 1\rm\ GHz$ provided that the number of particles with Lorentz 
factor $\sim 30$ (e.g. between 20 and 40) is of order $10^{24}$. We also 
note that FRB spectrum depends on the Lorentz factor distribution $\partial
N/\partial \gamma$. The radiation from particles with too large Lorentz 
factors will be beamed away from the observer line of sight, so we
expect that there should be a sharp break in the FRB spectrum somewhere above
$10\rm\ GHz$ (but not too far above 10 GHz, since the chance that an observer 
is within the beaming cone of particles with larger Lorentz factors
decreases as $\gamma^{-2}$).

\begin{figure}
  \centering
\includegraphics[width = 0.45 \textwidth,
  height=0.35\textheight]{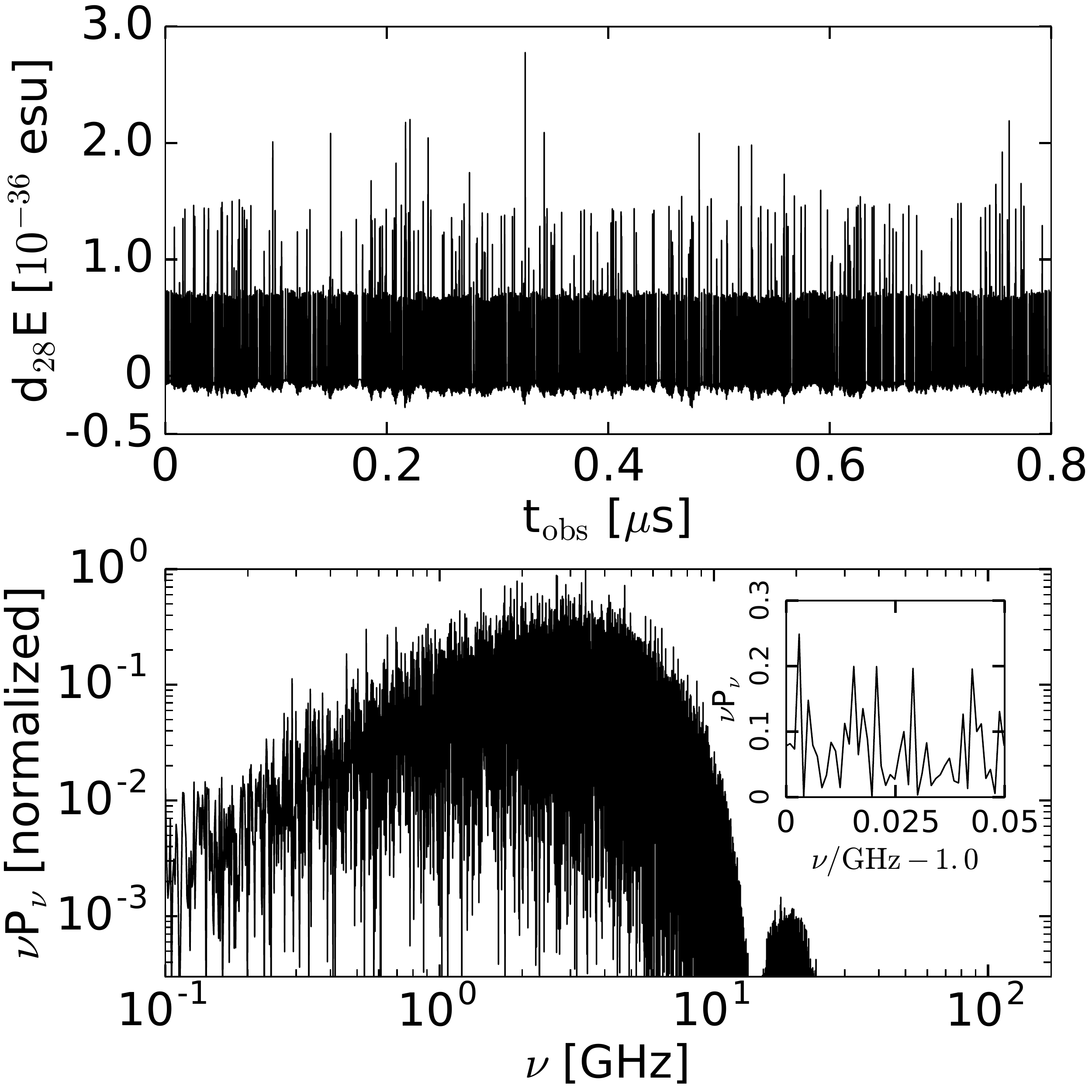}
\caption{The emission from 2000 identical patches of the case
  with $\Delta\theta = 0.3/\gamma^3$ and $\Delta \alpha = 1.0/\gamma$
  in Fig. (\ref{fig:coh_vol}). Adjacent patches are separated by 
  $\delta T$, and $\delta T$ is taken to be in Poisson distribution
  with mean value $5\times10^{-10}\rm\ s$.
}\label{fig:ran_pat}
\end{figure}

The observed FRB duration is affected by the DM smearing and broadening 
due to scatterings in the ISM/IGM. If the intrinsic duration of an FRB 
is $\delta t_{\rm in}$ (which is determined by the duration of magnetic 
reconnection), the number of coherent patches needed is $\sim 
\delta t_{\rm in}/1\rm\ ns\sim 10^6 \delta t_{\rm in, -3}$; it should be
pointed out that we receive radiation from only one of these patches
at a time, and the million patches are part of a semi-continuous 
outflow in a current sheet which we have broken up into a large number
of causally connected segments for the ease of calculation.  As a simple 
demonstration, we show in
Fig. (\ref{fig:ran_pat}) the emission from 
2000 identical patches that turn on and off on a timescale of $\delta T$
(in observer frame); $\delta T$ is taken to have the Poisson distribution 
with mean value $5\times10^{-10}\rm\ s$. The spectrum has strong fluctuations
in frequency intervals as small as $\sim$ MHz (Fig. \ref{fig:ran_pat}).
 To model the observed FRB
spectrum with rich fine-scale structure, one may need to add up
contributions from different patches in more realistic ways. For
example, particle flow could be episodic on $\mu$s--ms timescales and
different patches could have different polarizations due to the
rotation of the NS or non-stationary B-field configuration. 
Scintillation and plasma lensing probably also play an important role in 
shaping the observed spectrum \citep[e.g.][]{2016Sci...354.1249R}.

The EM emission directly coming from the source region is linearly
polarized, with electric field perpendicular to the primary B-field
and the line of sight. Due to the rotation of the NS, and the 
magnetic field orientation in the current sheet changing on $\mu$s
timescale, the polarization angle is likely to be time dependent. 
The polarization state of photons can also change substantially as they
propagate through plasma along the line of sight in the NS magnetosphere, 
magnetar wind nebula, and possibly supernova remnant. One would need
to take these effects into consideration in order to model and interpret 
the polarization properties of FRB radiation.

\subsubsection{Two stream instability in presence of electric field for 
relativistic relative velocity \& its effect on coherent radiation}

Since electrons and protons (or positrons) are accelerated by the
nearly static electric field parallel to $\vec B_0$ in opposite
directions, the two stream instability could have an effect on the
coherent radiation generation. We show that this
instability could lead to the formation of particle
bunches with longitudinal size on the order of $\sim \lambda$ on
timescales much shorter than $\gamma^2\nu^{-1}\sim 1\rm\ \mu s$. We
consider an electron-proton plasma, which is cold so that the state of
the particles transverse to the B-field is in the lowest Landau level; the 
analysis here is applicable to electron-positron plasma as well.

Our choice of frame is such that protons are at rest in the
  absence of a wave, which simplifies the algebra, but does not restrict 
linear stability analysis. The notation in this subsection is 
different from the rest of the paper. We denote particle density in the 
fluid comoving frame without a prime (in previous sections we had used prime), 
and all remaining quantities are measured in the rest frame of the proton 
and are also denoted without a prime. Variables in the lab frame 
will be denoted by the subscript ``lab''.

The electron and proton unperturbed densities in their comoving frames 
are $n_e$ \& $n_p$, and the perturbations are $\delta n_e$ \& $\delta 
n_p$ respectively. The 4-velocities for electrons and protons (as 
measured in the comoving frame of unperturbed proton flow) are taken 
to be $u_e + \delta u_e$ and $u_p + \delta u_p$; $u_p = (1,0,0,0)$. The 
perturbed equations for particle flux conservation follows from 
$(n u^\mu)_{,\mu} = 0$, and are
\begin{equation}
   \delta n_{p,t} + n_p \delta u_{p,z}^z = 0
    \label{dnp}
\end{equation}
\begin{equation}
   n_e \delta u_{e,t}^t + \delta n_{e,t} u_e^t + (\delta n_e u_e^z + n_e
     \delta u_e^z)_{,z} = 0
    \label{ne-pert1}
\end{equation}
It follows from $u_p^\mu u_{p\mu} = -1$ and $u_p^z=0$ (since perturbations 
are being carried out in the rest frame of protons) that $\delta u_{p}^t = 0$,
 but $\delta u_{p}^z\not=0$. Moreover, perturbation of the equation
 $u_e^\mu u_{e\mu} = -1$, yields
\begin{equation}
     u_e^t \delta u_e^t = u_e^z \delta u_e^z.
\end{equation}
We use this relation to eliminate $\delta u_e^t$ in equation (\ref{ne-pert1})
\begin{equation}
    \delta n_{e,t} + v\delta n_{e,z} + (n_e v/c\gamma) \delta u_{e,t}^z 
      + (c n_e/\gamma) \delta u_{e,z}^z = 0,
 \label{dne}
\end{equation}
where $\gamma \equiv u_e^t$, and $v/c \equiv u_e^z/u_e^t$.

The momentum equation for the cold plasma is:
$T^{\mu\nu}_{,\nu} = q F^{\mu\nu} u_\nu - F^\mu_{rad}$ or $u^\nu u^\mu_{,\nu} 
= (q/m) F^{\mu\nu} u_\nu - F^\mu_{rad}/m$; where $T^{\mu\nu} = m n 
u^\mu u^\nu$ is the energy momentum tensor for cold plasma, $F^{\mu\nu}$ 
is the electro-magnetic second rank tensor, $F^\mu_{rad}$ is the radiation 
reaction force, and $q$ and $m$ are particle charge and mass. 
The z-component of the momentum equation is
\begin{equation}
    u^z_{,t} + (u^z/u^t) u^z_{,z} = {q\over m} E_z - {F^z_{rad}\over m}
    \label{u_pz}
\end{equation}
The electric field $E_z$ is in the proton/positron comoving frame,
but since particles are streaming in the $z-$direction, the $z$-component
of electric field in the lab and electron comoving frames are also $E_z$.
The electric force on a charged particle is balanced by the radiation 
reaction force, $q E_z - F^z_{rad} = 0$, for the unperturbed system.
The perturbation to $F^z_{rad}$ is a non-local quantity (because it
depends on all the particles in the patch that are radiating coherently), 
and is not easy to include in the stability analysis of the system\footnote{
The effect of the radiation reaction force is likely to stabilize the
system. An increase in $q E_z$, causes particles to be accelerated 
to a higher Lorentz factor, which then causes a higher radiative power output
-- curvature radiation power scales as $\gamma^4$ (equation \ref{pe}) -- and the
resulting higher radiation reaction force tries to restore the 
equilibrium state of the system.}. We ignore $\delta F^z_{rad}$ in the
perturbation analysis presented below, and due to this we might perhaps be
overestimating the importance of the two-stream instability.

The perturbation of the z-component of the momentum equation for protons 
and electrons are:
\begin{equation}
   \delta u_{p,t}^z = {q\over m_p c} \delta E_z,
    \label{du_pz}
\end{equation}
and
\begin{equation}
    \delta u_{e,t}^z + v \delta u_{e,z}^z = -{q \delta E_z\over m_e c}.
    \label{du_ez}
\end{equation}
Finally, it follows from the Maxwell equations that
\begin{equation}
  \delta E_{z,z} = 4\pi q \left[ \delta n_p - \gamma \delta n_e - v n_e 
      \delta u_{e}^z/c \right].
    \label{dEz}
\end{equation}
Note that the density perturbation $\delta n_e$ is defined in the
 electrons' comoving frame, so we have the $\gamma$ factor on the
 right-hand side because the equation is in the proton/positron rest frame.

Equations (\ref{dnp}), (\ref{dne}), (\ref{du_pz}), (\ref{du_ez}) and
 (\ref{dEz}) are five equations in five variables that we solve to determine
the dispersion relation and two-stream instability growth time. 
Taking all perturbed variables to be proportional to $\exp(i\omega t - i kz)$,
these five equations can be combined and reduced to the following 
two equations 
\begin{equation}
   -{i m_e c^2\gamma(\omega - vk)^2 \over
       v\omega - k c^2} {\delta n_e\over n_e} + q \delta E_z = 0
    \label{dne-dEz1}
\end{equation}
and
\begin{equation}
    n_e\left[{v (\omega - vk) \over v\omega - k c^2} - 1\right]
     {\delta n_e\over n_e} + ik\left[ 1 - { \omega_p^2\over 
        \omega^2}\right]  {\delta E_z\over 4\pi q \gamma} = 0
    \label{dne-dEz2}
\end{equation}
We finally obtain the dispersion relation, which is:
\begin{equation}
 (\omega - v k)^2 \left[ 1 - {\omega_p^2\over \omega^2}\right] - {\omega_e^2
    \over \gamma^2} = 0, 
    \label{disperse}
\end{equation}
where
\begin{equation}
    \omega_p^2 \equiv  {4\pi q^2 n_p\over m_p}, \quad\quad{\rm and} 
        \quad \quad \omega_e^2 \equiv  {4\pi q^2 n_e\over m_e}.
    \label{plasma-w}
\end{equation}
The electron plasma frequency is
\begin{equation}
  \omega_e  \sim 
   (2\times10^{12}\,{\rm rad\,s}^{-1}) n_{e,15}^{1/2}
    \label{plasma_w1}
\end{equation}
It should be emphasized that $k$ and $\omega$ are wavenumber 
and complex frequency in proton/positron rest frame; $v$ and $\gamma$ are 
the unperturbed velocity and Lorentz factor of the electron fluid in 
the rest frame of proton/positron, and $n_e$ and $n_p$ are particle 
number densities in the rest frame of electron and proton respectively.

\begin{figure}
\centerline{\hbox{\includegraphics[width=9cm, height=13cm, angle=0]{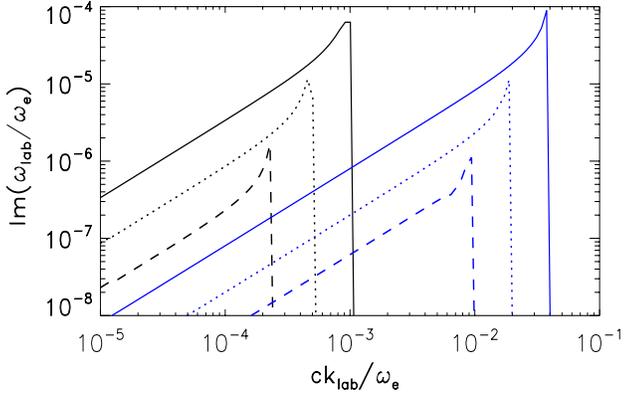}}}
\vskip -7.3cm
\caption{Shown here is the growth rate of two-stream instability in the
lab frame --- imaginary part of $\omega_{lab}/\omega_e$ --- for 
an electron-proton plasma with 
$\gamma_{_{e,lab}}=\gamma_{_{p,lab}} = 25$, 50 and 100 (black solid, dotted and dashed lines respectively)  as a function of wavenumber in the lab frame;
$\gamma_{_{e,lab}}$ \& $\gamma_{_{p,lab}}$ are Lorentz factors of electrons and
protons in the lab frame; we note that the Lorentz factor is independent  
of particle mass when acceleration is balanced by the radiation reaction. 
There are no unstable modes for the e-p plasma for $\gamma_{_{e,lab}}=25$ 
when $ck_{lab}/\omega_e \gae 10^{-3}$.  Also shown is the imaginary part of 
$\omega_{lab}/\omega_e$ for electron-positron plasma with $\gamma_{_{e,lab}}=
\gamma_{_{p,lab}} = 25$, 50 and 100 (blue solid, dotted and dashed lines 
respectively). 
}
\label{Figure-1}
\end{figure}

It can be shown from equation (\ref{disperse}) that two-stream instability 
is present when $[1+(m_e\gamma^2/m_p)^{1/3}]^{3/2}/(v\gamma k/\omega_e)<1$,
i.e. for sufficiently small wavenumber the system is unstable to bunching
up of particles. The growth rate of instability, Im$(\omega$), is obtained 
by finding the two complex
roots of equation (\ref{disperse}), which are complex conjugates of each other.
The growth rate for $e^-$--$p^+$ and $e^-$--$e^+$ plasma for several 
different values of $\gamma$ are shown in Fig. (\ref{Figure-1}). 
Since $\gamma$ in equation (\ref{disperse}) is the electron Lorentz 
factor as measured in the rest frame of protons, it is 
equal to $2\gamma_{_{p,lab}} \gamma_{_{e,lab}}$, where $\gamma_{_{p,lab}}$ and 
$\gamma_{_{e,lab}}$ are Lorentz factors of protons and electrons measured 
in the lab frame. 

Electron clumps of longitudinal size $\sim \lambda\sim 30$ cm are required
for the coherent curvature radiation model considered in this work. 
The electron density needed to generate a typical FRB luminosity of 
$10^{43}$ erg s$^{-1}$ is $n_e\sim 10^{16}$ cm$^{-3}$ (see eq. \ref{Liso2} 
and the discussion following it --- but note that $n_e$ in this section is
same as $n_e'$ in that equation). Therefore, $\omega_e \sim 5\times10^{12}$ 
rad s$^{-1}$ (eq. \ref{plasma_w1}). Hence, for a $e^-$--$p^+$ plasma of
$\gamma_{_{e,lab}}\sim 30$ (needed for the FRB curvature radiation, as per
eq. \ref{gam}), the clump size ($\sim 2\pi/k_{lab}$, for the fastest growing 
mode) turns out to be $\sim 50$ cm, and the growth time for the instability 
is about 30 ns in the lab frame (see Fig. \ref{Figure-1}); the time required
for the formation of a clump is longer (due to causality limit), and is
of order $0.5\,\mu$s. For a 
$e^-$--$e^+$ plasma, clumps of size $\sim 30$ cm also form on a timescale of
$\sim 0.5$ $\mu$s in the lab frame. Since the emission
from one coherent patch lasts for 1 ns in the observer frame, which
corresponds to $\gamma_{_{e,lab}}^2\sim 10^3$ ns in the lab frame,
particle clumps form on a time scale smaller than that required for 
the coherent curvature radiation to operate. 

We note that the electric field associated with the clumping of
electrons is much weaker than $E_\parallel$ estimated in equation
(\ref{E_parallel1}) and the electric field associated with the FRB EM
wave energy $E_{\rm \perp EM}$ in equation (\ref{E_em}). Even in the
extreme case where electrons are 
fully separated with protons/positrons in different clumps with
longitudinal separation $\lambda$, the electric field due to clumping
is roughly given by $E_{\rm \parallel, clump}\sim 4\pi N_eq/(\gamma_{_{\rm e,
  lab}}^2\lambda^2)$, where 
$N_e$ is the total number of particles in one clump and $\gamma_{_{\rm
  e, lab}}^2\lambda^2$ is the transverse area of a coherent patch. For
parameters required to produce FRB luminosities, $N_e \sim 10^{24}$,
$\gamma_{_{\rm e, lab}}\sim 30$ and $\lambda\sim 30\rm\ cm$, we obtain
$E_{\rm \parallel, clump} \sim 5\times10^9\rm\ esu$. In reality, the
clumps are only partially charged and the electric field due to
clumping is much smaller than this value. Therefore, the electric field 
associated with clumping will not be able to disperse the clumps; the
back reaction of radiation on the clump can not disperse them either,
because this reaction force is balanced by the reconnection 
electric field $E_\parallel$ (eq. \ref{E_parallel1}).  
Moreover, the electric field of charged clumps is not
sufficiently strong to convert plasma waves to the EM waves at FRB
luminosities.

\subsubsection{Perturbation of particle trajectories by EM field and
  particle collisions --- coherence survives}

We showed in \S\ref{sec-particle} that particle velocity
distribution along the B-field need not be narrow for generation of
coherent radiation. However, even a small dispersion in particle velocity 
perpendicular to the B-field can ruin the coherence. Also, the primary
B-field lines along which particles in the source region are moving should be
nearly parallel to each other (within an angle of $\gamma^{-1}$).

We show here that despite the presence of strong perpendicular
electric and magnetic fields in the source region ($B_{ind} \sim
10^{12}$ G; equation \ref{B_ind1}), 
and collision between particles with opposite charges moving relative to each 
other at close to the speed of light, the distribution of particle velocity
perpendicular to the B-field $\vec B_0$ is unaffected. The reason
for this comes down to the fact that the energy of excited Landau levels are
so large that particles stay in the ground state in spite of the large, 
time dependent, fields. And therefore, the coherence is preserved.

The energy of relativistic electrons in a magnetized medium
is (see Appendix A)
\begin{equation}
  \epsilon(p_z, n)^2 = m_e^2 c^4 + p_z^2 c^2 + 2 \hbar\omega_c(n+1/2) m_e c^2,
    \label{E-landau}
\end{equation}
where $p_z$ is the component of electron's momentum along the B-field 
line, $n$ is a positive integer, and
\begin{equation}
    \omega_c = {q B_0\over m_e c}
    \label{wc}
\end{equation}
is the electron cyclotron frequency. In the limit $p_z/(m_e c) \approx 
\gamma \gg 1$, the expression for electron energy can be approximated 
as
\begin{equation}  \epsilon(p_z, n) \approx p_z c + \hbar\omega_c(n+1/2)/\gamma,
    \label{E-landau1}
\end{equation}
Thus, the energy of the electron corresponding to the lowest Landau
level (for $n=0$) is 
\begin{equation}
     \epsilon(p_z, 0) \approx p_z c + 1.1\times10^{-2} m_e c^2 B_{0,14}/
             \gamma_2
    \label{E-landau2}
\end{equation}
The plasma temperature in the direction transverse to the B-field should 
be cold ($T \lae 10$ keV) -- even though electrons are streaming along magnetic 
field lines with a Lorentz factor $\sim10^2$ -- otherwise higher Landau 
levels would be populated and the emergent radiation cannot be coherent. 

The motion of electrons perpendicular to the B-field is damped due to
the cyclotron radiation on a very short time scale of 
$\sim (10^{-17}$ s) $B_{0,14}^{-2}\gamma_2$ 
(in the lab frame). So the electron velocity dispersion perpendicular
to the B-field (which is proportional to the transverse temperature)
should be very small unless there is a mechanism or instability that operates
on a much shorter timescale and excites electrons to higher Landau levels. 

From equation (\ref{E-landau}) we see that the effective electron momentum
transverse to the B-field, in the first excited level ($n=1)$, is
$p_\perp = (3\hbar\omega_c m_e)^{1/2}$. This is much larger than the 
transverse momentum kick given to the electrons, $\sim q^2 n_e^{1/3}/c$ 
(classically), when electrons and protons undergo Coulomb collision in 
the FRB source. 
Particle collisions, therefore, cannot dislodge electrons from their ground 
Landau state. 

The transverse electric field associated with FRB radiation
is so strong ($E_{\perp EM} \sim 10^{11}$ esu; equation \ref{E_em}) that
it shifts the energy of the Landau levels by an amount that is much larger 
than $\hbar\omega_c/\gamma$. However, it can be shown that this shift
is exactly the same for all Landau levels, and since $\omega_c/\gamma$
is larger than the EM wave frequency ($\sim 1$ GHz) by a factor
$\sim 10^9$, electrons adapt to this periodic shift in the energy levels
adiabatically and remain in the ground state, i.e. the strong EM field 
associated with the radiation does not kill the coherence. 

The B-field produced by charged particles in the source region streaming along
$\vec B_0$ is also strong ($B_{ind}\sim 10^{12}$ G; equation \ref{B_ind1}),
and its direction changes on a length scale smaller than $\lambda$. However,
electrons adapt to this changing B-field adiabatically since the
de Broglie wavelength of electrons in transverse direction, 
$(2\hbar/\omega_c m_e)^{1/2}\sim 3\times10^{-9} B_{0,14}^{-1/2}\rm\
cm$ (see Appendix A), is much smaller than the 
curvature radius of the B-field. 

\subsection{Propagation of radio waves through supernova remnant}

Waves generated near the surface of a relatively young NS travel
through the supernova remnant left behind in the explosion that
produced the NS. We show in this sub-section that the contribution to
DM from the supernova remnant is nearly constant for a period of a few  
years as required by the data for FRB 121102 from which 17 bursts 
have been detected \citep{2016Natur.531..202S, 2016ApJ...833..177S}; this point has already
been made recently by \citep{2016MNRAS.461.1498M, 2016ApJ...824L..32P, 
2017arXiv170102370M}. We also show that the optical depth
of the supernova remnant for free-free absorption is small. 

Let us consider that 10$M_\odot$ of material is ejected in a supernova 
that produced the NS, and the initial speed of the remnant is 
10$^9$ cm s$^{-1}$. The particle density of the medium in the vicinity 
of the explosion is taken to be 10 cm$^{-3}$. In this case the deceleration 
radius for the remnant is $\sim 7\times10^{18}$ cm, and the deceleration 
time is about 200 years. The mean particle density of the ejecta $t$ yrs 
after the explosion is $\sim 30 t_2^{-3}$ cm$^{-3}$,
and the electron column density is $\lae 10^{20} t_2^{-2}$ cm$^{-2}$, if 
the ejecta is significantly ionized as suggested by
\citet{2017arXiv170102370M}; 
where $t_2 = t/(10^2$ yrs). Thus, the 
contribution of the ejecta to the FRB DM is $\sim 35 t_2^{-2}$ pc cm$^{-3}$. 
And the change to DM in three years is $\sim 2 t_2^{-3}$ pc cm$^{-3}$.

Seventeen radio bursts were detected from FRB 121102 in a period of 3
years, and the DM values for these events were the same (about 557 pc cm$^{-3}$) during this 
period within the error of measurements. Thus, the age of the NS for 
this FRB cannot be smaller than 50 years. A similar lower limit on the 
age is obtained from the consideration that the DM from the host galaxy
(at $z=0.19$) does not exceed $\sim 10^2$ pc cm$^{-3}$
\citep{2017ApJ...834L...7T}. 

The free-free absorption optical depth for the remnant at 1 GHz is
$\tau_{ff}\sim 3\times 10^{-2} n_e^2 t_2 T^{-3/2}\sim 3\times10^{-5} t_2^{-5} 
T_4^{-3/2}$; where $T$ is the temperature of the remnant. 
\citet{2017ApJ...834L...7T} find $\tau_{ff}\lae 10^{-3}$ for FRB 121102, 
and that too suggests the NS age to be $\gae$50 yrs.

It can be inferred from the large B-field strength of ($\gtrsim 10^{14}$
G) that the NS is most likely a magnetar. The empirical age of
Galactic magnetars is $\sim10^3$ to $10^4\rm\ yr$
\citep[e.g.][]{2013MNRAS.434..123V}, which may be taken as the 
upper limit of the age of the FRB progenitor.

\bigskip
{\it To summarize the main result of \S3, the picture that emerges 
is that FRB radiation is produced in a patch of comoving size $\gamma\lambda$, 
where magnetic reconnection provides the necessary electric field to 
accelerate charged particles to Lorentz factor $\sim 30$, and these particles
produce coherent curvature radiation. The observed FRB radiation can be
produced in a small region of transverse size $\sim 10^3$ cm (i.e.
$\eta \sim 1$), with B-field strength of $\gtrsim 10^{14}$ G, and particle 
density of the order Goldreich-Julian density. 
The reconnection of the B-field is sustained for $\sim$ 1 ms, and 
the electric field in the current sheet parallel to the primary
B-field is $\sim10^{11}$ esu. }

\section{Predictions of the Model}
\label{predict}

We provide in this sub-section a number of predictions of the model we
have proposed for FRBs. In essence, the model is that radio waves
are produced coherently in the magnetosphere of a NS with strong B-fields 
that undergo forced reconnection due to perhaps the movement of the 
neutron star crust (where the fields are anchored) or emergence of 
sub-surface flux tubes above the NS surface.
The distortion of the B-field lines in the crust or magnetosphere 
builds up over a period of time, until it reaches a critical state
and becomes unstable and field lines are reconfigured, dissipating
a fraction of the B-field energy in a current sheet in the process. 
This is a well known process and is responsible for
the energy release in solar flares
\citep[e.g.][]{2011LRSP....8....6S}.

We begin by writing down the coherent curvature luminosity and the
strength of the electric field parallel to the B-field in a nearly model
independent form, and then specialize to magnetic reconnection to express
length scales in terms of parameters appropriate for a current sheet.

\begin{equation}
     L_{iso} \approx {8\pi^2\over 3} q^2 c (n_e')^2 \ell_t^4 \gamma^4,
    \label{Liso4}
\end{equation}
which is essentially the same as equation (\ref{Liso1}), except that we have 
taken $N_{patch} = 1$ (as discussed in \S\ref{curvature-rad}),
the transverse area of the source is $\ell_t^2$, and the expression
for curvature radiation frequency (equation \ref{nu_R}) has been used to
eliminate the wavelength $\lambda$. The electric field strength
($E_\parallel$) is 
calculated by balancing the acceleration due to electric field with the
radiation back reaction force (as was done in \S3 leading to equation 
\ref{E_parallel}) and we re-express that result as:
\begin{equation}
   E_\parallel \sim {4\pi q n_e' \ell_t^2 \gamma^2\over 3 \rho} \sim
   \left[{L_{iso} 
   \over c \rho^2}\right]^{1/2},
    \label{E_parallel3}
\end{equation}
where as before $\rho$ is the radius of curvature of the B-field.

Thus far we have not made any reference to reconnection and current
sheets, 
and so the above results are broadly applicable. Next, we specialize to
current sheets, and make a few predictions. Since we have a rather limited
first-principle understanding of the properties of current sheets and
particle acceleration, the following discussion is mostly qualitative.

The thickness of the current sheet is likely to be related to the plasma 
frequency ($\omega_p$), and accordingly we take $\ell_t \sim 
\eta_p c/\omega_p$; where $\eta_p$ is a dimensionless parameter which 
could be of order $10^2$--10$^3$ as suggested by numerical simulations
\citep[e.g.][]{2016MNRAS.462...48S}. Substituting for $\ell_t$
in equation (\ref{E_parallel3}) we find 
\begin{equation}
   \gamma \sim \left({3 q E_\parallel \rho\over \eta_p^2 m_e
       c^2}\right)^{1/2} \propto
   \eta_{p}^{-1}E_{\parallel}^{1/2}\rho^{1/2},
\label{gam_cs}
\end{equation}
which is independent of the plasma density ($n_e'$). Using equation 
(\ref{gam_cs}) for $\gamma$, we calculate the frequency at which 
particles radiate most of their energy
\begin{equation}
   \nu \sim {\gamma^3 c\over 2\pi\rho} \sim {c\over 2\pi \rho} \left[ {3 q
      E_\parallel \rho \over \eta_p^2 m_e c^2}\right]^{3/2} \propto \eta_{p}^{-3}
      E_\parallel^{3/2} \rho^{1/2},
\label{nu_peak}
\end{equation}
and
\begin{equation}
 L_{iso} \sim E_\parallel^2 \rho^2 c \propto \eta_{p}^{4} \nu^{4/3}\rho^{4/3}.
   \label{Liso-nup}
\end{equation}

The electric field $E_\parallel$, $B_0$, $\theta_B$ (the angle between 
field lines on the opposite sides of the current sheet), and $\eta_p$
likely vary substantially from one reconnection event to another. 
Therefore, several general predictions can be made for FRBs according
to the model described in this work.
\begin{enumerate}
%\begin{enumerate}[(i)]
\item Electrons are likely accelerated to very different Lorentz factors
    in different bursts or different regions in one burst. Hence, in
    addition to the FRBs observed at a few  
GHz frequency, the model predicts that there are other FRBs that radiate
at frequencies much higher than GHz (since $\nu\propto \gamma^3$);
some FRBs may radiate most of the energy at 10 GHz whereas others
might be at 
$\sim10^{14}$ Hz. Since $\nu\propto E_\parallel^{3/2}$, it is unlikely to
find FRBs at frequencies much larger than $\sim10^{14}$ Hz; this is
because $\nu \sim 10^{14}$ Hz requires $E_\parallel\sim 10^{14}$ esu --- which is
of order the Schwinger limit on electric field strength of 
$m_e^2 c^3/(q^2 \hbar)$ --- and fields of larger strength cannot be 
sustained for long as they lead to spontaneous creation of electron-positron 
pairs and breakdown of the vacuum. High energy photons can be produced due to
a process such as the resonant inverse-Compton scattering if the conditions
in the source region are just right, but that requires fine tuning of 
parameters.
\item FRB event rate should decrease with increasing frequency
 roughly as\footnote{One caveat is that the variation of FRB rate with 
  frequency (for a given flux limit) depends on particles' Lorentz factor
  distribution $\partial N/\partial \gamma$, and the combined angular size of
  all the beams in case there are multiple beams associated with a source.
  The FRB rate $\propto \nu^{-2/3}$ should
  be taken as a rough estimate.} $\gamma^{-2}\propto \nu^{-2/3}$, or 
  perhaps more steeply, because of possible
  enhancement to the beaming associated with the transverse source size.
\item The intrinsic durations of FRBs at frequencies much larger than GHz
   may be of the same order as GHz-FRBs provided that there are a large 
   number of independent beams (each of angular size $\gamma^{-1}$) that 
   contribute to the total duration. On the other hand, if the radiation 
   is produced by a single beam sweeping across the observer line of 
   sight --- which is the most conservative possibility for these transients ---
  then the observed duration at higher frequencies would be smaller than 
   a few ms since the higher frequency radiation is produced by electrons 
   with larger Lorentz factors.

\item The luminosity function of FRBs in a fixed frequency band should be 
     broad since $L_{\rm iso}\propto \eta_p^{4}$. This might be 
     the reason for the large flux variations (by at least an order of
     magnitude) observed for the different outbursts of the repeating 
     FRB 121102 \citep{2016Natur.531..202S, 2016ApJ...833..177S}.
\end{enumerate}

The intrinsic coherent radiation luminosity decreases with decreasing 
frequency, and the observed pulse width increases rapidly at lower 
frequencies due to wave scattering \citep[roughly as $\nu^{-4}$, but
see][]{2016ApJ...832..199X} by turbulence 
in the ISM/IGM and DM smearing ($\nu^{-3}$). These effects cause
the observed flux to decrease sharply  
at lower frequencies. To make matters worse, lower frequency waves may
have 
difficulty in getting out of the supernova ejecta as a result of 
free-free absorption. So there is likely to be a low frequency cut-off
to the FRB observations which might be of order a few hundred MHz.

\section{Summary \& discussion}

The very high brightness temperature of FRBs suggest that the radiation
process is coherent, which means that the comoving source size cannot
be larger than the comoving-frame wavelength $\lambda'$ of the
radiation we observe. Thus, the electric field 
associated with radiation at the source is of order 10$^{11}$ esu (and 
the B-field is 10$^{11}$ G). Furthermore we show that the curvature 
radiation can account for the observed FRB properties without any need 
for fine tuning. The current associated with charged particles' motion
along field lines produces transverse B-field (with roughly 
cylindrical shaped field lines) that is sufficiently strong to perturb 
particle trajectories and destroy coherence unless the primary field is
$\gae 10^{14}$ G (see eq. \ref{B0-lim}). The strong B-field also ensures 
that electrons stay in the ground state of Landau levels in spite of 
several strong perturbations that are present in the source region 
that otherwise would excite particles to higher Landau level and destroy 
coherence. A particle density of $\sim 10^{17}$ cm$^{-3}$ (in lab frame)
is sufficient to account for the observed isotropic equivalent FRB
luminosity 10$^{43}$ erg s$^{-1}$; the total number of charged
particles in one coherent patch with Lorentz factors $\gamma \sim 30$
is of order 10$^{24}$; a coherent patch produces radiation for $\sim
1$ ns, and therefore  
the total number of particles responsible for a typical FRB transient 
is $\sim 10^{30}$, and the mass of this matter is about a kg for an 
electron-positron plasma.

However, because the particles are radiating in phase, their
  radiative cooling time is extremely small --- of order 10$^{-15}$ s
  or six orders of magnitude smaller than the wave period --- unless
  there is a powerful acceleration mechanism that balances the
  radiative losses and maintains  
the particle speed. An electric field of strength $\sim 10^{11}$ esu, 
parallel to the primary B-field, is the required mechanism to sustain 
the particle motion for at least the lab-frame
travel time ($\gamma^2\nu^{-1}\sim\rm \mu s$) over which GHz EM waves
are produced. Such an electric field could be produced during forced 
magnetic reconnection near the surface of a magnetar. This
could also  
explain why we see the bursts repeat episodically. The total energy release 
in these bursts, corrected for beaming, is estimated to be of order 
$10^{36}$ ergs, whereas the total energy in the B-field is at 
least $\sim 10^{45}B_{0,14}^2$ ergs.

The intrinsic durations of some FRBs may be much longer than the
light-crossing time of a NS. If this is true, the process that drives
the magnetic reconnection should be relatively ``slow''. One possible
scenario is that B-field flux emerges from 
below the NS surface due to buoyancy
\citep[e.g.][]{1995ApJ...440L..77M, 2012MNRAS.425.2487V} and then
reconnects with pre-existing B-field in the magnetosphere. Another 
possibility is the slow movement of the NS crust where the fields are
anchored. It is currently unclear what this process might be.

According to the model we have described, the Lorentz factor of electrons
(and hence the peak frequency of the spectrum), and the isotropic
luminosity are dependent mainly on the thickness of the current
sheet $\ell_t=\eta_pc/\omega_p$ ($\omega_p$ being the plasma
frequency) and the strength of the electric field parallel to the B-field,
$E_\parallel$. These two parameters could have large variations among
different bursts or at different locations of one burst.
Thus, according to the model we have described, the luminosity
function for FRBs should be very broad. Also, the mechanism we have
described should produce short duration bursts at frequencies much
larger than 1 GHz\footnote{Waves at frequencies much smaller than 1 GHz
 may suffer from free-free absorption by the surrounding medium
 (e.g. supernova remnant) and much more severe scattering
 broadening.} (up to $\sim 10^{14}$ Hz). The rate of
bursts, however, is predicted to decrease with increasing peak
frequency because higher frequency photons require larger Lorentz
factor of particles or smaller curvature radius of B-field. The former is a
much stronger effect, and the burst rate is expected to
decrease with frequency at least as $\nu^{-2/3}$.

There are several issues that we have not discussed in detail, which 
future works should address. In particular, studying the propagation of
polarized waves through the NS magnetosphere and surrounding medium in
the host galaxy may help our understanding of the odd spectrum of FRBs
and their polarization properties. The distortion of B-field lines, and
reconnection for an extremely large magnetization parameter plasma
($\sigma \gae 10^{15}$) are also topics that require separate
investigations.

\section{acknowledgments}
We thank M. Bailes, D. Lorimer, E. Petroff, M. Lyutikov,
A. Spitkovsky, A. Piro, J. Cordes and A. Stebbins for
discussions of FRB observations and possible 
emission mechanisms, B. Zhang for reading the manuscript and for his 
excellent comments, and F. Guo and S. Paban for useful discussions
regarding magnetic reconnection and quantum effects in strong magnetic 
field respectively. We are highly indebted to the referee, Jonathan Katz,
for his careful reading of the manuscript and for his numerous excellent
comments and suggestions that substantially improved the paper and 
clarified a number of conceptual issues.
WL was funded by the Named Continuing Fellowship at the University of
Texas at Austin.

%%%%%%%%%%%%%%%%%%%%Begin the Reference%%%%%%%%%%%%%%%%%%%%%%%%%%

%%%%%%%%%%%%%%%%%%%%End the Reference%%%%%%%%%%%%%%%%%%%%%%%%%%%

\vfill\eject

\appendix{}
\section{Landau levels for relativistic particles}

We calculate the energy states of a charged particle in a strong B-field.
Particle speed along the field line is highly relativistic, and the field is
very strong so that the particle spin is aligned with the field. We, therefore,
ignore particle spin, and consider Klein-Gordan equation with magnetic
potential (the electric field is taken to be zero):
\begin{equation}
   \left(i\hbar\vec\nabla - {q \vec A\over c}\right)^2 \psi + m^2 c^2 \psi
      = -{\hbar^2\over c^2} {\partial^2\psi\over \partial t^2},
\end{equation}
or
\begin{equation}
   \hbar^2\nabla^2\psi + {2i \hbar q\over c} \vec A\cdot\vec\nabla\psi
    + {i\hbar q\over c}(\vec\nabla\cdot\vec A)\psi - {q^2 A^2\over c^2}\psi
    - m^2 c^2 \psi = {\hbar^2\over c^2} {\partial^2\psi\over \partial t^2}.
\end{equation}

Let us consider a uniform B-field, $\vec B = B_0 \hat z$,
and vector potential $\vec A = -y B_0 \hat x$ corresponding to it. The wave 
function has a non-trivial dependence on the $y$-coordinate, and we
express it as
\begin{equation}
   \psi(\vec x, t) = u(y) \exp(ip_z z/\hbar + i p_x x/\hbar - i E t/\hbar),
\end{equation}
and substitute that in the Klein-Gordan equation to obtain
\begin{equation}
    {d^2u\over d y_1^2} + {u\over \hbar^2 c^2}\left[ E_1^2 - q^2 B_0^2 y_1^2
        \right] = 0,
\end{equation}
where
\begin{equation}
     y_1 \equiv y - {cp_x\over q B_0}, \quad{\rm and}\quad E_1^2 \equiv
       E^2 - p_z^2c^2 - m^2 c^4.
\end{equation}

The equation for $u$ is that of a harmonic oscillator, and therefore
the quantum states have energies as follows:
\begin{equation}
  E_1^2 = 2qB_0 c\hbar(n + 1/2), 
\end{equation}
or
\begin{equation}
 E^2 = m^2 c^4 + p_z^2 c^2
        + 2 \hbar \omega_c(n + 1/2) m c^2,
\end{equation}
where
\begin{equation}
   \omega_c = {q B_0\over m c}
\end{equation}
is the cyclotron frequency.

The wave function $u_n$, for the Landau state $n$, is given by:
\begin{equation}
 u_n(y_1) = {1\over \sqrt{2^n n!}} \left[ {2m \omega_c\over h}\right]^{1/4}
      e^{-{m\omega_c y_1^2\over 2 \hbar}} H_n(y_1 \sqrt{m\omega_c/\hbar}),
\end{equation}
where $H_n$ is the Hermite polynomial of n-th order
\begin{equation}
   H_n(x) = (-1)^n e^{x^2} {d^n\over dx^n} e^{-x^2}.
\end{equation}
The spatial extent of the wave-function in the ground state is
\begin{equation}
   \lambda_{DB\perp} \sim \sqrt{ {2\hbar\over m_e \omega_c}}
\end{equation}
This is a result that we use in \S3.

\end{document}